\documentstyle[aps,prd,eqsecnum,twocolumn,epsf]{revtex}

\newcommand{\square}{\kern1pt\vbox{\hrule height 1.2pt\hbox{\vrule
width 1.2pt\hskip 3pt
\vbox{\vskip 6pt}\hskip 3pt\vrule width 0.6pt}\hrule
height 0.6pt}\kern1pt}
\newcommand{\beq}{\begin{equation}}
\newcommand{\beqn}{\begin{eqnarray}}
\newcommand{\eeq}{\end{equation}}
\newcommand{\eeqn}{\end{eqnarray}}

\begin{document}

\draft

\twocolumn[\hsize\textwidth\columnwidth\hsize\csname
@twocolumnfalse\endcsname

\title{Dynamics of Scalar field in a Brane World}
\author{Shuntaro Mizuno$^1$, Kei-ichi Maeda$^{1,2,3}$, and Kohta
Yamamoto$^4$\\~}
\address{$^1$Department of Physics, Waseda University, Okubo 3-4-1,
Shinjuku, Tokyo 169-8555, Japan\\[-1em]~}
\address{$^2$ Advanced Research Institute for Science and Engineering,
Waseda University, Shinjuku, Tokyo 169-8555, Japan\\[-1em]~}
\address{$^3$ Waseda Institute for Astrophysics,
Waseda University, Shinjuku, Tokyo 169-8555, Japan\\[-1em]~}
\address{$^4$ Shinano Mainichi Shimbun Inc.,
Minami-agatamachi 657, Nagano 380-0836, Japan\\[-.7em]~}

\date{\today}
\maketitle
\begin{abstract}
We study the dynamics of a scalar field  in the brane cosmology. We
assume that  a scalar field is confined in our 4-dimensional world. As
for  the potential  of the  scalar field, we discuss three typical
models: (1) a power-law potential,  (2) an inverse-power-law
potential, and (3) an exponential potential. We show that the
behavior of the scalar field is very different from a conventional
cosmology when the energy density square term is dominated.

\end{abstract}
\vskip 1pc 

\vskip 1pc
]

\section{Introduction}                           %

A scalar field has played a very important role in cosmology. The
problems  of the standard Big-Bang theory, such as the horizon,
and flatness problems
could be resolved by the inflationary scenario\cite{inflation}.  In
this scenario, the universe expands quasi-exponentially because of the
vacuum energy, whose origin in
many models are given by a scalar field (inflaton). Among inflationary
scenarios, the chaotic inflation, which has a power-law type
potential $( V(\phi) \sim  \phi^n : n$   is an integer) is the
simplest model. Another example in which, a scalar field plays an
important role  is a cosmological phase transition and formation of
topological defects\cite{defect}. The observed dark
energy could be also explained by a scalar field whose energy density
decreases in time (quintessence)\cite{Ratra_Peebles,Quintessence}.
In a typical quintessence model, the potential of the
scalar field is assumed to be  an inverse-power-law ($V(\phi) \sim
\phi^{-\alpha}$ : $[\alpha>0])$.
An  exponential-type ($ V(\phi) \sim
\exp[-\lambda \phi]$) is naturally expected in many supergravity
theories\cite{SUGRA}. Hence those scalar fields  are extensively
studied by many  authors in the conventional Einstein gravity theory.
However, recent higher dimensional unification scenario based on a brane
world suggests that  the gravitational law could be  different from
      the   Einstein's one at the early stage of the universe
or at high-energy scale.

In the brane-world scenario, which
is based on a superstring or M-theory, our
universe is embedded in higher dimensions. Standard-model  particles
are confined in this four-dimensional hypersurfaces (3-branes), while
gravity  propagates in a higher-dimensional bulk
space\cite{3-brane,String,H-W,RS}. Among them, the Randall-Sundrum's
second model is very interesting because it provides a new type of
compactification of gravity. Assuming a 3-brane with a positive tension
is embedded in 5-dimensional anti de-Sitter bulk spacetime, the
conventional four-dimensional gravity theory is recovered in low energy
limit, even though the extra dimension is not compact\cite{RS}. While, at
high energy scale, gravity   differs from  the conventional Einstein
theory\cite{SMS}. Many authors  discussed the geometrical aspects of
the brane and its dynamics (For a review, see, \cite{Brev}),  as well as
its cosmological implications\cite{Bcmgy1,Bcmgy2,Bcmgy3}.

Based on the Randall-Sundrum's
second model, if we assume that our brane is homogeneous and
isotropic (Friedmann-Robertson-Walker universe),
the difference from
conventional cosmology appears as two new terms in the Friedmann equation,
i.e. the  quadratic term of energy-momentum and   dark
radiation\cite{Bcmgy1,SMS}. As for the dark radiation,  its
effect is constrained by the nucleosynthesis.  It is also  diluted away in
the inflationary era. Hence, the important change in  scalar field
dynamics, which we are going to discuss,
may be due to the appearance of the term of  quadratic energy density .

The purpose of the present paper is to study dynamics of a scalar
field confined on the brane, taking into account the effect of the
quadratic-energy-density term.
We discuss three typical potentials appeared in many cosmological
implications; a power-law potential,  an inverse-power-law
potential, and  an exponential potential.
The analysis of the quintessential potential
will complete our previous discussion\cite{maeda,mizuno_maeda}.

In Sec. \ref{sec.2}, we present the basic equations assuming the
Randall-Sundrum's
second  model. Next, in Sec. \ref{sec.3}, we show
cosmological solutions in both  scalar-field dominant  and
radiation dominant eras.  These solutions are found analytically by
using some approximations.

In Sec.
\ref{sec.4}, we first study  linear perturbations of the solutions, and
then  in Sec.
\ref{sec.5}, we analyze those global stability. Sec. \ref{sec.6} is
devoted to summary and discussion.

        \section{BASIC EQUATIONS}
        \label{sec.2}

We analyze dynamics of a scalar field in the Randall-Sundrum's second
brane scenario\cite{RS}, because the model is simple and concrete. It
is,  however,  worthwhile noting that the present result may be also
valid in other types of brane world models, in which a quadratic term of
energy-momentum tensor generically appears. In the brane world, all
matter fields and forces except gravity are confined on the 3-brane in a
5-dimensional spacetime. The
extra-dimension is not compactified, but gravity is confined  in the
brane, resulting in the Newtonian gravity in our world.
Since the gravity is confined in the brane, it is described by the
intrinsic metric of the 4-dimensional  brane spacetime. By use of Israel's
junction condition and assuming
$Z_2$  symmetry, the gravitational equations on the 3-brane
is given by\cite{SMS}
\beqn
^{(4)}G_{\mu\nu}=-^{(4)}\Lambda g_{\mu\nu}+\kappa_4^2
T_{\mu\nu}+\kappa_5^4 \Pi_{\mu\nu}-E_{\mu\nu},
\label{geq}
\eeqn
where $^{(4)}G_{\mu\nu}$ is the Einstein tensor with respect
to the intrinsic metric $g_{\mu\nu}$, $^{(4)}\Lambda$ is the 
F
4-dimensional cosmological constant, $T_{\mu\nu}$
represents the energy-momentum tensor of matter fields confined on the
brane  and
$\Pi_{\mu\nu}$ is defined by its quadratic form. $E_{\mu\nu}$ is a part
of the 5-dimensional Weyl tensor and carries some information about a bulk
geometry.
$\kappa_4^{\;2}=8\pi G_4$ and
$\kappa_5^{\;2}=8\pi G_5$ are 4-dimensional and 5-dimensional
gravitational constants,  respectively. In what follows, we use the
4-dimensional Planck mass $m_4 \equiv \kappa_4^{~-1}
=(2.4\times10^{18}{\rm GeV})$ and the 5-dimensional
Planck mass $m_5\equiv
\kappa_5^{~-2/3}$, which could be much smaller than  $m_4$.

Assuming the Friedmann-Robertson-Walker
spacetime in our brane world,  we find the effective Friedmann equations
from Eq.(\ref{geq}) as
\beqn
H^2+{k\over
a^2} &=&\frac{1}{3}{}^{(4)}\Lambda +\frac{1}{3m_4^{\;2}}
\rho+\frac{1}{36m_5^{\;6}} \rho^2+\frac{{\cal C}}{a^4}
\label{fr1}
\\
\dot{H}-{k\over a^2}&=&-\frac{1}{2m_4^{\;2}} \left(P+\rho\right)
-\frac{1}{12m_5^{\;6}} \rho\left(P+\rho\right)-\frac{2{\cal C}}{a^4}
\label{fr2}
\eeqn
where $a$ is a scale factor of the Universe, $H=\dot{a}/a$ is its Hubble
parameter, $k$ is a curvature constant,
$P$ and
$\rho$ are the total pressure and total energy density of matter fields,
respectively.
${\cal C}$ is a constant,  which term describes "dark"  radiation
coming from $E_{\mu\nu}$\cite{Bcmgy1}. In what follows, we consider only
the flat Friedmann model ($k=0$) and  assume that $^{(4)}\Lambda $
vanishes  for simplicity.

As for matter fields on the brane, we consider  a scalar field $\phi$
as well as the conventional radiation fluid, i.e.
$\rho=\rho_{\phi}+\rho_{\rm r}$, where $\rho_{\phi}$, and
$\rho_{\rm r}$ are the energy densities of scalar field $\phi$,
and of radiation fluid.
We consider only the early stage of the universe, at which the quadratic
term is dominant and then  matter fluid can be ignored.

Although a 5-dimensional scalar field living in the
bulk\cite{maeda_wands} may also appear in a brane world scenario, in
this paper we
      only consider  a 4-dimensional scalar field confined on the brane.
The origin of such a scalar field might be found as a result of
condensation of matter fields  confined on the brane such as fermions.

Since the energy and momentum of each field on the brane  are conserved in
the present model, we find the dynamical equation for  a
scalar field as a conventional one, i.e.
\beqn
\ddot{\phi}+3H\dot{\phi}+\frac{dV}{d\phi}=0,
\label{sde}
\eeqn
where $V$ is a potential of the scalar field.
The energy density of
the scalar field is
\beqn
\rho_\phi=\frac{1}{2}\dot{\phi}^2 + V(\phi) .
\label{sed}
\eeqn
For the energy density of radiation fluid, we have
\beqn
\dot{\rho_{\rm r}}+4H\rho_{\rm r}&=&0,
\label{ecr}
\eeqn
which is integrated as $\rho_{\rm r} \propto a^{-4}$.

As the Universe expands, the energy density  decreases.
This means that the quadratic term was very important in the early stage
of the Universe. Comparing two terms (the conventional energy density
term and the quadratic one), we find that the quadratic term dominates
when
\beq
\rho > \rho_c \equiv 12 m_5^{~6}/m_4^{~2} .
\eeq

When the quadratic term is dominant,
the expansion law of the Universe is modified.
For example, the expansion law in radiation dominant era is
$a\propto t^{1/4}$ instead of the conventional one $t^{1/2}$.
Since we know the behavior of the scalar field in the linear term dominant
stage, which is the conventional cosmological model, in this paper we
study   the  behavior of the scalar field only  in the quadratic term
dominant stage.

The nucleosynthesis gives a constraint on fundamental constants, if the
quadratic-energy-density dominant stage exists. The nucleosynthesis must
take place in the conventional radiation dominant era to explain the
amount of light elements. Hence, assuming that  the energy density at
$a=a_c$ is  dominated by radiation as $\rho_c \sim   (\pi^2/30) g T_c^4$,
where $g$ is the degree of freedom of particles,
the temperature of the universe
$T_c$ must be higher than that of nucleosynthesis, i.e. $Tc> T_{NS} \sim
1  \rm{MeV}$. This constraints implies
$m_5 >1.6 \times 10^4(g/100)^{1/6}
(T_{\rm NS}/1\:\:{\rm MeV})^{2/3} \:\:{\rm GeV}$.

        \section{Cosmological solutions}
\label{sec.3}
There are two interesting limiting cases: one is the scalar-field
dominant  era and the other is the radiation dominant era.
We discuss those solutions separately.

        \subsection{Solutions in scalar-field dominant era}
Assuming the scalar field dominance, we find  the Friedmann equation
(\ref{fr1}) as
\beqn
H=\frac{1}{6m_5^{\;3}} \left[ \frac{1}{2}\dot{\phi}^2 +
V(\phi) \right].
\label{frqsde}
\eeqn

We  consider the following three types of the potential:
$V(\phi)  =\mu^{\alpha+4}\phi^{-\alpha},
\mu^4\exp[-\lambda
\frac{\phi}{m_4}]$ and $\frac{1}{2}m^2\phi^2$,
$(\frac{1}{4}\lambda \phi^4)$ in order.

\vspace*{0.3cm}
{\bf A.1}~~$ V(\phi)  =\mu^{\alpha+4}\phi^{-\alpha}$
\vspace*{0.3cm}

The equation for the scalar field (\ref{sde}) is now
\beqn
\ddot{\phi}+3H\dot{\phi}-\alpha\mu^{\alpha + 4}\phi^{-\alpha-1}=0.
\label{ivsdesd}
\eeqn
Inserting Eq. (\ref{frqsde}) into Eq.
(\ref{ivsdesd}), we find a second order differential equation for $\phi$.
It is easily shown that the asymptotic behaviors of the solutions are
        classified into three cases:  (a) slow rolling (a
potential term dominant) solution $\left( \alpha < 2 \right)$,  (b)  a
solution in which the potential term balances with the  kinetic term
$\left( \alpha = 2 \right)$, and  (c) a kinetic-term dominant solution
$\left( \alpha > 2
\right)$.

Assuming a slow rolling condition, we find that
\beqn
H \approx \frac{1}{6m_5^{\;3}} \mu^{\alpha+4}\phi^{-\alpha} ~{\rm and} ~
\dot{\phi} \approx \frac{\alpha}{3H} \mu^{\alpha+4} \phi^{-\alpha-1},
\eeqn
which lead to $\phi\dot{\phi} =2\alpha m_5^{\;3}$.
Then  we find the solution
\beqn
\frac{\phi}{m_5}=2\left[ \alpha m_5
\left(t-t_0\right)\right]^{1/2},
\label{sevsr}
\eeqn
where $t_0$ is an integration constant. In order to keep the slow-rolling
condition, we obtain $\alpha < 2 $ because
$\ddot{\phi}~(\sim t^{-3/2}) \ll V'(\sim t^{-(\alpha+1)/2})$ and
$\dot{\phi}^2(\sim t^{-1}) \ll V(\sim t^{-\alpha/2})$.
The universe expands as
\beqn
a = a_0 \exp\left\{ \left[H_0  \left(t-t_0 \right)\right]^{(2-\alpha
)/2}\right\},
\label{sqrtif}
\eeqn
where a constant $H_0$ is given by
\beq
{ H_0\over m_5}=
\left[3\left(2-\alpha\right)
\left(2\sqrt{\alpha}\right)^\alpha\right]^{-{2\over(2-\alpha)}}
\left( {\mu\over
m_5}\right)^{2(\alpha+4)\over(2-\alpha)}.
\eeq
This solution describes an inflationary evolution, whose
expansion rate is slower than that of the conventional exponential
inflation, but faster than the power-law type. It may be interesting to
discuss a spectrum of density perturbations for such an inflationary
scenario.

For $\alpha = 2$, we find a power-law solution\cite{maeda} as
\beqn
a = a_0 (t/t_0)^p,
\label{plif}
\eeqn
with
\beqn
p=\frac{1}{6}\left[1+\frac{1}{8}\left(\frac{\mu}{m_5}\right)^6\right].
\label{plindex}
\eeqn
and
\beqn
{\phi\over m_5}=2[2m_5t]^{1/2}.
\label{sevpl}
\eeqn
The scalar field energy density evolves as
\beqn
\rho_{\phi} \propto t^{-1} \propto a^{-1/p}.
\label{sfedqrple}
\eeqn

If $p>1$, i.e. $\mu > 40^{1/6}m_5 \approx 1.85 m_5$,
we have a power-law inflationary solution\cite{footnote1}, which is an
attractor of the  present  system as we will see later.
While, if $p < 1/4$, i.e. $\mu < 4^{1/6}m_5 \approx 1.26 m_5$,
this solution is no longer an attractor, leading to the radiation
dominant era (see Sec. \ref{sec.4}).

If $\alpha > 2$, a kinetic-term dominant solution gives
the asymptotic behavior of the present system
(see Sec. \ref{sec.4}).
If  a kinetic term is  dominant, we find
\beqn
H\approx
\frac{1}{12m_5^{\;3}}\dot{\phi}^2 ~{\rm and}~
\ddot{\phi}+3H\dot{\phi}\approx 0,
\eeqn
      leading to
$\ddot{\phi} \approx -(1/4m_5^{\;3})\dot{\phi}^3$,
which is integrated as
\beqn
\frac{\phi}{m_5}=\pm 2\sqrt{2}m_5^{1/2}\left(t-t_0\right)^{1/2}
+\frac{\phi_0}{m_5},
\label{sevkd}
\eeqn
where $t_0$ and $\phi_0$ are integration constants.
For the solution with
$+$ sign,  the potential term decreases as $V  \propto
t^{-\alpha/2}$, while the kinetic  term drops as $\dot{\phi}^2
        \propto t^{-1}$, then the kinetic term is always dominant if
$\alpha > 2$.   In this case, we find that the Universe expands as
\beqn
a = a_0 (t/t_0)^{1/6},
\label{sckd}
\eeqn
which is exactly the expansion law of the quadratic-term dominant
universe with stiff matter.

On the other hand, for the solution with $-$ sign, $\phi
\rightarrow
0$ as $t \rightarrow t_0$, and then $V
\rightarrow
\infty$. Then before reaching $t_0$ , the potential term becomes
dominant. However, since the potential-dominant solution is not any
attractor, it turns out that the former kinetic-dominant solution
is eventually reached (see
Sec. \ref{sec.4}). Consequently, the kinetic-term dominance is the asymptotic
behavior for
$\alpha > 2$.  This guarantees the radiation, if it exists, eventually
dominates the scalar-field, which makes
initial conditions for a successful quintessence
wider\cite{maeda,mizuno_maeda}.

\vspace*{0.3cm}
{\bf A.2}~~$
V(\phi)=\mu^4\exp[-\lambda\frac{\phi}{m_5}]$
\vspace*{0.3cm}

The exponential potential is steeper than the inverse-power-law potential.
Then the asymptotic solution is the same as Eqs. (\ref{sevkd}) and
(\ref{sckd}) in the case (1) with
$\alpha > 2$.
Then the radiation dominant era is eventually  realized.

\vspace*{0.3cm}
{\bf A.3}~~ Chaotic inflationary potential
\vspace*{0.3cm}

The inflationary solution of this model was
already obtained\cite{maart_et.al}, for completeness, we list here the
analytic solution including its
oscillating phase.

For power-law potential such as $V={1\over 2}m^2\phi^2$ or ${1\over
4}\lambda \phi^4$, we find a stronger inflation  than the
conventional one\cite{maart_et.al}.\\
(i) For the model
$V(\phi)=\frac{1}{2}m^2\phi^2$, assuming a slow rolling condition, we
find that
\beqn
H \approx {m^2\over 12m_5^3}  \phi^2 ~{\rm and}~
\dot{\phi} \approx -{m^2\over 3H} \phi,
\eeqn
which lead to $\phi \dot{\phi}=-4m_5^{\;3}$. We then obtain
the  solution
\beqn
\phi=\pm 2\sqrt{2}m_5^{\;3/2}(t_0-t)^{1/2},
\label{sevplpi}
\eeqn
\beqn
a=a_0 \exp\left[-\frac{m^2}{3}(t_0-t)^2\right],
\label{scplpi}
\eeqn
where $t_0$ and $a_0$ are integration constants.

The slow-rolling condition is broken when $| \ddot{\phi} | \sim | V' |$
or
$(1/2)\dot{\phi}^2 \sim V$. Both conditions give the time when inflation
ends, i.e.
$t_f \sim t_0 - (2m)^{-1}$.
In order for inflation to take place in the quadratic energy density
dominant  stage, $\rho_\phi (\phi_f) > \rho_c$, which implies
$m > 3m_5^{\;3}/m_4^{\;2}$.

After the end of inflation, there appears  an oscillation period,
which solution is given approximately as
\beqn
\phi= \frac{2m_5^{\;3/2}}{mt^{1/2}} \sin mt,
\label{sevplpo}
\eeqn
\beqn
a = a_0 (t/t_0)^{1/3}.
\label{scplpo}
\eeqn
The scale factor evolves just as the case with a pressureless perfect
fluid (dust fluid) at the quadratic term dominant stage. Since the
amplitude of
$\phi$ decreases more slowly than the conventional case because of
the small expansion rate, we expect a sufficient particle production in
the preheating phase\cite{tsujikawa}.

(ii) For the model with  $V(\phi) = \frac{1}{4} \lambda \phi^4$,
      the behavior of the scalar field in  the inflationary phase is not so
different from the case of the previous massive inflaton model.  Assuming
a slow rolling  condition, we find that
\beqn
H \approx {\lambda\over 24 m_5^{\;3}}
\phi^4 ~{\rm and}~
\dot{\phi} \approx -{\lambda\over 3H}\phi^3,
\eeqn
      which
lead to $\phi \dot{\phi} = -8 m_5^{\;3}$. We obtain the
solution
\beqn
\phi=\pm 4 m_5^{\;3/2}(t_0-t)^{1/2},
\label{sevgplpi}
\eeqn
\beqn
a=a_0 \exp\left[-\frac{32 \lambda m_5^3}
{9}(t_0-t)^{3}\right],
\label{scgplpi}
\eeqn
where $t_0$ and $a_0$ are integration constants.

The slow-rolling condition is broken at $t=t_f \sim t_0- 2^{1/3}
/(\lambda^{1/3} 4 m_5)$ when $(1/2)\dot{\phi}^2 \sim V$.
Another condition ($| \ddot{\phi} | \sim | V' |$)
corresponds to the time
$t_f^{~\prime}
\sim t_0 - 1/(\lambda^{1/3} 4 m_5)$, which appears after
$t_f$.  The  scale factor at the end of inflation $a_f \equiv a(t_f) = a_0
\exp[-1/9] \sim  a_0$. In order inflation to take place in the quadratic
energy density dominant  stage, $\rho_\phi (\phi_f) > \rho_c$, which
implies
$\lambda > (27/32)(m_5/m_4)^{\;6}$.

For the oscillating period, it is shown that there is a time-averaged
relation  between the potential term and the kinetic term of the scalar
field, i.e.
$\langle \dot{\phi}^2 \rangle = \lambda \langle \phi^4 \rangle$ and
then the  scalar field behaves  approximately as the radiation
fluid\cite{Reheating}. Then we find that
\beqn
a\propto t^{1/4}~~{\rm and}~~\langle
{\phi} \rangle\propto t^{-1/4} .
\label{rad_kin}
\eeqn
Introducing a conformal
time
$\eta \equiv \int dt a(t)^{-1}$
and the conformal field
$\varphi \equiv a \phi$,  the equation  for $\varphi$
(Eq. (\ref{sde})) is
\beqn
\varphi'' + \lambda \varphi^3 -\frac{a''}{a} \varphi = 0,
\label{sdevarphi}
\eeqn
where $'$ stands for the derivative with respect to the conformal time.
Since
the time averaged value of $\varphi$ is almost constant  while
$(a''/a) \sim \eta^{-2}$, the third term can be ignored.
In the conventional cosmology, since $a \sim t^{1/2} \sim
\eta$, the  third term  exactly vanishes.  Then
we can integrate Eq.(\ref{sdevarphi}), finding the
elliptic function as
\beqn
\varphi = \tilde{\varphi}~ {\rm cn} \left(x-x_0,
\frac{1}{\sqrt{2}}\right),
\label{varphi}
\eeqn
where a constant $\tilde{\varphi}$ is an oscillation amplitude of
$\varphi$, $x \equiv \sqrt{\lambda}\tilde{\varphi} \eta$ is a
dimensionless conformal time variable, and $x_0$ is the value of $x$
when the inflaton field starts to oscillate coherently.
This solution is the same  as that  obtained in the
conventional cosmology. However  the corresponding time evolution of the
time-averaged scalar field $\langle \phi \rangle$ is given as
$\langle \phi \rangle \propto a^{-1} \propto t^{-1/4}$, which  decreases
more slowly than the conventional one.  We
study  this effect at the preheating stage in \cite{tsujikawa}.


        \subsection{Solutions in radiation dominant era}

      In the radiation dominant era, the scale factor
expands as $ a \propto t^{1/4}$. We have to solve the equation for
a scalar field.

The basic equation is now
\beqn
\ddot{\phi}+\frac{3}{4t}\dot{\phi}+V'=0.
\label{sderde}
\eeqn
We shall consider again the same potentials discussed in the
previous section in order.

\vspace*{0.3cm}
{\bf B.1}~~$V(\phi)  =\mu^{\alpha+4}\phi^{-\alpha}$
\vspace*{0.3cm}

The equation of motion for the scalar field
(\ref{sde}) is now
\beqn
\ddot{\phi}+\frac{3}{4t}\dot{\phi}-\alpha\mu^{\alpha + 4}\phi^{-\alpha-1}
=0.
\label{sderdeiv}
\eeqn
We find an analytic solution for $\alpha<6$\cite{maeda}, that is
\beqn
\phi &=&\phi_0 \left( {t\over t_0}\right)^{\frac{2}{\alpha + 2}},
\label{sevkpc}\\
&& {\rm with}~~~
\phi_0^{\alpha+2}={2\alpha(\alpha+2)^2 \over 6-\alpha}\mu^{\alpha+4}t_0^2,
\label{phi0}
\eeqn
where $t_0$ and $\phi_0 (>0)$ are integration constants.
Eq. (\ref{phi0}) requires $\alpha<6$.

The energy density of the scalar field evolves as
\onecolumn
\beqn
\rho_{\phi} &=&
{3\alpha(\alpha+2) \over
6-\alpha}V_0\left({t\over t_0}\right)^{-\frac{2\alpha}{\alpha + 2}}
\nonumber\\
& =& {3\alpha(\alpha+2) \over
6-\alpha}V_0\left({a\over a_0}\right)^{-\frac{8\alpha}{\alpha + 2}},
\label{sfedqrrde}
\eeqn
where $V_0=V(\phi_0)$ and $a_0=a(t_0)$.

The density parameter of the scalar field, which we denote as
$\Omega_\phi$, is

\beq
\Omega_\phi = {\rho_\phi \over \rho_{\rm r} +\rho_\phi} \approx
{\rho_\phi \over \rho_{\rm r}} = \Omega_\phi^{(0)} \left({a\over
a_0}\right)^{\frac{4(2-\alpha)}{\alpha + 2}},
\eeq
where $ \Omega_\phi^{(0)}={3\alpha(\alpha+2) \over
6-\alpha}V_0/\rho_{\rm r}(t_0) $ is an initial value of the density
parameter.

If $\alpha > 2$, $\Omega_\phi$ decreases with time,
just contrary to the tracking solution in the conventional quintessence
model.  The scalar field energy decreases faster than that of the
radiation. This is a new interesting feature  because the "initial"
smallness of  a  quintessence-field energy  could be dynamically
explained\cite{maeda,mizuno_maeda}.
      If $ \alpha = 2$, $\Omega_\phi$ is
constant until the liner term becomes  dominant. This is the so-called
"scaling" solution. The scalar field energy drops at the same rate as
that of the radiation. On the other hand, if $ \alpha < 2$, $\Omega_\phi$
increases, and then  the  universe get into a scalar field energy
dominant era discussed above, which is an inflationary solution.

For $\alpha>6$, we find a kinetic-term dominant solution
\beq
\phi=\phi_0\left({t\over t_0}\right)^{1/4},
\label{rad_kin2}
\eeq
where $\phi_0 (>0)$, and $t_0$ are integration constants, decided by the
initial value only. In this case, they do not include the characteristic
scale, because there is no contribution of the potential term for
the energy density of the scalar field.

\vspace*{0.3cm}
{\bf B.2}~~$V(\phi)=\mu^4\exp[-\lambda\frac{\phi}{m_5}]$
\vspace*{0.3cm}

This case is not so interesting, because the kinetic energy
dominant solution (\ref{rad_kin2})
gives an asymptotic behavior just as the same as the case of
scalar field dominant era.

\vspace*{0.3cm}
{\bf B.3}~~Chaotic inflationary potential
\vspace*{0.3cm}

(i) For the model $V(\phi)=\frac{1}{2}m^2\phi^2$,
the solution is given approximately as
\beqn
\phi=\phi_0 \left({t\over t_0}\right)^{-3/8} \sin mt .
\label{sevplprd}
\eeqn
The energy density of the scalar field $\rho_\phi$, which  evolves
approximately as
$\rho_\phi \propto t^{-3/4} \propto a^{-3}$, is the same as
that of dust fluid. then the scalar field energy will eventually dominate
unless its energy is transferred into radiation via a reheating process.

(ii) For the model $V(\phi)=\frac{1}{4} \lambda \phi^4$,
the behavior is the same as Eq.(\ref{varphi}),
because  the time evolution of the scale factor is the same.

We summarize the above solutions in the next table.
\vskip 0.5cm

\begin{tabular}{|@{}r   @{}c || @{}c|@{}c|@{}c|}
\hline
       ~~~~potential     &        & S~&&~R~($a\propto t^{1/4}$)\\
\hline\hline
    $|$ & $ \alpha < 2$
            & ~inflation : Eqs.(\ref{sqrtif}),(\ref{sevsr})
          &$\leftarrow$ &
\\
\cline{2-4}
     $|$ &   $ \alpha =2$
          & ~power-law :  Eqs.(\ref{plif}),(\ref{sevpl})
          &$\times$& ~power-law : Eq.(\ref{sevkpc})
\\
\cline{2-4}
\raisebox{1.5ex}[0pt]{$\phi^{-\alpha}$} ~~$|$  & $~2 < \alpha < 6$&
     &$\rightarrow$&
\\
\cline{2-2}\cline{4-5}
$|$  &  $~\alpha\geq 6$  & ~kinetic dominance:  Eqs.(\ref{sckd}),(\ref{sevkd})
&&\\
\cline{1-2}
$~e^{-\lambda\phi}$&&
&\raisebox{1.5ex}[0pt]{$\rightarrow$}&~\raisebox{1.5ex}[0pt]{kinetic
dominance:Eq.(\ref{rad_kin2})}\\
\hline
${1\over 2}m^2\phi^2$&&~inflation:Eqs.(\ref{scplpi}),(\ref{sevplpi})
$\rightarrow$
oscillation:Eqs.(\ref{scplpo}),(\ref{sevplpo})&$\leftarrow$&oscillation:Eq.(
\ref{sevplprd})\\
\hline
${1\over 4}\lambda\phi^4$&&~inflation:Eqs.(\ref{scgplpi}),(\ref{sevgplpi})
$\rightarrow$
oscillation:Eqs.(\ref{rad_kin}),(\ref{varphi})&$\times$&oscillation:Eq.(\ref
{varphi})\\
\hline
\end{tabular}
\begin{table}[h]
\caption[table1]{\footnotesize{The asymptotic behavior in
quadratic-energy-density dominated stage. S and R denote scalar-field  dominant
era ($\rho_\phi
\gg
\rho_{\rm r}$) and   radiation dominant era ($\rho_{\rm r} \gg \rho_\phi $),
respectively. The arrows describe the final fate.}}
\vskip .2cm
\end{table}

\twocolumn
\section{Stability analysis}
\label{sec.4}

In the previous section, we have listed up several analytic solutions in
the quadratic energy density dominated era, which are summarized in Table 1.
If  those solutions are one of the attractors in the dynamical system,
such spacetimes may be naturally realized in the history of the universe.
Therefore, it is very important to study their stability.

In this section,  we study the stability of  the  solutions
    against linear perturbations.
Although we may need the detailed analysis for the solutions for the
chaotic inflation
    and the following oscillation,
we expect that such solutions are found for a wide range of initial
conditions just as the same as conventional chaotic inflationary scenario.
Hence  we focus only on the stability of solutions of inverse-power-law
and exponential potentials.

Since the cosmological solutions are time dependent, we first have to find
new variables by which the solutions are described as fixed points in the
dynamical system.

        \subsection{scalar field dominated stage}

\subsubsection{the power-law solution}

First, we investigate the power-law solution
(\ref{plif}),(\ref{sevpl}) with an inverse-power-law potential
with $\alpha=2$, i.e. $V=\mu^6\phi^{-2}$.
The asymptotic behavior of  the solution
is
\beqn
{H\over m_5}&\sim&  {p\over (m_5t)},
\\
{V\over m_5^{4}}&\sim&{{\cal M}_0^{~2}\over
(m_5t)},\\
{\dot{\phi}\over m_5^2}&\sim&{\sqrt{2}
\over (m_5t)^{1/ 2}}
\eeqn
where ${\cal M}_0\equiv  \left({\mu/
m_5}\right)^{3}/2\sqrt{2} $ and $p=(1+{\cal M}_0^{~2})/6$.

For these time dependence, we introduce new variables as
\beqn
{\cal{H}} &\equiv& tH,\label{nv1}\\
{\cal{V}} &\equiv& m_5^{-3} t V ,\label{nv2}\\
{\cal{P}} &\equiv& m_5^{-3/2} t^{1/2}\dot{\phi}.
\label{nv3}
\eeqn

In terms of these variables, the basic equations are expressed as,
\beqn
{\cal{H}}' &=& {\cal{H}}(1-6{\cal{H}}+{\cal{V}}),\\
{\cal{V}}' &=& {\cal{V}}-(\sqrt{2}{\cal M}_0)^{-1}
{\cal{P}} {\cal{V}}^{3/2}, \\
{\cal{P}}' &=& \frac{{\cal{P}}}{2}(1-6{\cal{H}})+
(\sqrt{2}{\cal M}_0)^{-1}{\cal{V}}^{3/2},
\eeqn
with the following constraint,
\beqn
{\cal{H}}=\frac{1}{12}{\cal{P}}^2 + \frac{1}{6}{\cal{V}},
\label{nv4}
\eeqn
where a prime denotes a derivative with respect to new time coordinate
$\tau=\ln (m_5t)$.

These suggest that there are only two independent variables.

For the basic equations, the power-law solution is described
by a fixed point
$({\cal{H}}_0,{\cal{V}}_0,{\cal{P}}_0) =(p,{\cal M}_0^{~2},\sqrt{2}) $.

After eliminating ${\cal{H}}$ with Eq.(\ref{nv4}),
the perturbation equations around this solution are
\beqn
\delta{\cal{V}}'& =& -\frac{1}{2} \delta{\cal{V}}
-{{\cal M}_0^{~2}\over \sqrt{2}} \delta{\cal{P}},
\\
\delta{\cal{P}}'& =&
\frac{1}{2\sqrt{2}}\delta{\cal{V}}
-\left(1+\frac{1}{2}{\cal M}_0^{~2} \right)\delta{\cal{P}}.
\eeqn

Because these equations do not include the explicit
time-dependence, we can discuss the stability of the solution,
by the eigenvalues of these coupled equations.

They are
\beqn
\left(-1, -\frac{p}{2}\right),
\eeqn
which are all negative, and then this solution is stable
against linear perturbations.

        \subsubsection{the kinetic dominant solution}

Next, we analyze the stability of the kinetic dominant solution
(\ref{sevkd}) with an inverse-power-law potential with $\alpha>2$ and an
exponential potential.
First we discuss the case with the potential
$V(\phi)=\mu^{\alpha+4}\phi^{-\alpha}$.
The asymptotic behavior of  the solution
is
\beqn
{H\over m_5}&\sim&  {1\over
6(m_5t)},
\\
{V\over m_5^{4}}&\sim&{\cal V}_0 (m_5t)^{-\alpha/2},\\
{\dot{\phi}\over m_5^2}&\sim&{\sqrt{2}
\over (m_5t)^{1/ 2}},
\eeqn
where ${\cal V}_0 \equiv 8^{-\alpha/2}(\mu/m_5)^{\alpha+4}$.

For such time-dependence, the  variables ${\cal
H}$ and ${\cal P}$ are given by the same relations (\ref{nv1})
and (\ref{nv3}), but ${\cal V}$ is newly introduced as
\beq
{\cal V} \equiv m_5^{-4}V e^{{\alpha\over 2}\tau},
\eeq
using the time coordinate $\tau=\ln (m_5t)$.
Then we find the following basic equations as,
\beqn
{\cal{H}}' &=& {\cal{H}} (1-6{\cal{H}} +
{\cal{V}}e^{-{\alpha-2\over 2}\tau}),
\label{calhiv}
\\
{\cal{V}}' &=& {\alpha\over 2}{\cal{V}}\left[1-\frac{\cal{P}}{\sqrt{2}}
\left({{\cal{V}}\over
{\cal{V}}_0}\right)^{\frac{1}{\alpha}}\right],
\label{calviv}
\\
{\cal{P}}' &=& \frac{{\cal{P}}}{2}(1-6{\cal{H}})+
{\alpha \over 2\sqrt{2}} {\cal{V}}_0^{-{1\over \alpha}}
{\cal{V}}^{\frac{\alpha+1}{\alpha}} e^{-\frac{\alpha-2}{2} \tau},
\label{calpiv}
\eeqn
with the following constraint equation,
\beq
{\cal H} ={1\over 6}\left[{1\over 2}{\cal P}^2 +{\cal V}
e^{-{\alpha-2\over 2}\tau}\right].
\label{scalar_energy}
\eeq

 From Eqs.(\ref{calhiv})-(\ref{calpiv}) with (\ref{scalar_energy}), in the
limit of
$\tau\rightarrow \infty$, we find the fixed point
$({\cal H}, {\cal V}, {\cal P}) =(1/6,{\cal{V}}_0,\sqrt{2})$,
which corresponds to the kinetic-term
dominant solution.
Eliminating ${\cal{H}}$ with Eq.(\ref{scalar_energy}) and
expanding the above  basic equations around this
fixed point as
$({\cal{H}},{\cal{V}}, {\cal{P}}) = (1/6 + \delta{\cal{H}},
{\cal{V}}_0+\delta{\cal{V}},
\sqrt{2} + \delta{\cal{P}})$, we find that the perturbation equations
are
\beqn
\delta{\cal{V}}' &= & -\frac{1}{2}\delta{\cal{V}}
-\frac{\alpha{\cal{V}}_0}{2\sqrt{2}}\delta{\cal{P}},
\label{calvpriv}
\\
\delta{\cal{P}}' &=& (\alpha-2) \frac{{\cal{V}}_0}{2\sqrt{2}}
e^{-\frac{\alpha-2}{2}\tau}-\delta{\cal{P}}.
\eeqn

One may think that this system is easily analyzed because
the time-dependent part in these equations
decays with time.  If we estimate the eigenvalues for the asymptotic
equations, we find that those are -1/2 and -1, which suggests the system
is stable. However, since the equations have explicit  time-dependence,
we have to be careful to conclude it  Fortunately, in this case,
we can easily integrate the above perturbed equations directly as
\beqn
\delta{\cal{V}}&= &\delta{\cal V}_0
e^{-{\tau\over 2}} +\frac{\alpha {\cal V}_0}{\sqrt{2}}
\delta{\cal P}_0 e^{-\tau}\nonumber\\
&&-{\alpha(\alpha-2)\over 2(\alpha-3)(\alpha-4)}{\cal V}_0^2
e^{-\frac{\alpha-2}{2} \tau}\\
\delta{\cal{P}}&= & \delta{\cal{P}}_0e^{-\tau}
-{(\alpha-2)\over \sqrt{2}(\alpha-4)}{\cal
V}_0 e^{-\frac{\alpha-2}{2} \tau},
\eeqn
if $\alpha\neq 3, 4$,
\beqn
\delta{\cal{V}}&= &\frac{3{\cal V}_0}{\sqrt{2}}\delta{\cal P}_0e^{-\tau}
-{3\over 4}{\cal V}_0^2\tau e^{-\frac{1}{2} \tau}+\delta{\cal
V}_0e^{-\frac{1}{2} \tau}
\\
\delta{\cal{P}}&= & \delta{\cal{P}}_0e^{-\tau}
+{1\over \sqrt{2}}{\cal{V}}_0 e^{-\frac{1}{2} \tau},
\eeqn
for $\alpha=3$, and
\beqn
\delta{\cal{V}}&= &\delta{\cal V}_0 e^{-\frac{1}{2} \tau}+2\sqrt{2}
{\cal V}_0\delta{\cal P}_0 e^{-\tau}+2{\cal V}_0^2(\tau+2) e^{-\tau}
\\
\delta{\cal{P}}&= & \delta{\cal{P}}_0e^{-\tau}
+{1\over \sqrt{2}}{\cal{V}}_0 \tau e^{-\tau}
\eeqn
for $\alpha=4$, where $\delta{\cal{V}}_0$ and  $\delta{\cal{P}}_0$
are the integration constants.
These show that $\delta{\cal{V}},\delta{\cal{P}} \rightarrow 0$
in the limit of $\tau\rightarrow \infty$, which means that
the kinetic dominant solution is stable
against linear perturbations for
$V(\phi) = \mu^{4+\alpha} \phi^{-\alpha}$, with $\alpha > 2$.
Note that for the case of $2<\alpha<3$, the perturbations drop
more slowly than those expected just from the above eigenvalues.

Next, we consider the case with an exponential potential
$V= \mu^4 e^{-\lambda\phi/m_5}$.
The asymptotic behavior of the  solution
is
\beqn
{H\over m_5}&\sim&  {1\over
6(m_5t)},
\\
{\phi\over m_5}&\sim&2\sqrt{2}
   (m_5t)^{1/ 2},\\
{\dot{\phi}\over m_5^2}&\sim&{\sqrt{2}
\over (m_5t)^{1/ 2}}.
\eeqn
In this case, because the potential decreases quite fast as
$V\propto \exp(-2\sqrt{2}\lambda (m_5t)^{1/2})$, we adopt
new variables
\beq
{\cal F}\equiv m_5^{-3/2} \phi~ t^{-1/2}
\eeq
instead of ${\cal {V}}$.
The basic equations are given as
\beqn
{\cal{H}}'& =& {\cal{H}}- \frac{1}{24}{\cal{P}}^4-
\frac{1}{12}\left(\frac{\mu}{m_5}\right)^4
{\cal{P}}^2 \exp\left(\tau-\lambda e^{\tau/2}{\cal{F}}\right),
\\
{\cal{F}}' & =&-{1\over 2}{\cal{F}}+{\cal{P}},
\\
{\cal{P}}' & =& \frac{{\cal{P}}}{2}(1-6{\cal{H}})+
\lambda \frac{\mu^4}{m_5^4} \exp\left({3\over
2}\tau-\lambda e^{\tau/2} {\cal{F}}\right),
\eeqn
with the following constraint equation,
\beq
{\cal H}={1\over 12}{\cal P}^2 +{\mu^4\over
6m_5^4}~\exp
\left(\tau-\lambda e^{\tau/2}{\cal {F}}\right).
\label{scalar_energy_exp}
\eeq

If ${\cal F} >0$, in the limit of $\tau\rightarrow \infty$, we find a
fixed point
$({\cal{H}}_0,{\cal{F}}_0, {\cal{P}}_0)= (1/6,2\sqrt{2},\sqrt{2})$, which
corresponds to the kinetic-term dominant
solution (\ref{sevkd}),(\ref{sckd}).

Expanding the basic
equations around this fixed point as
$({\cal{H}},{\cal{F}}, {\cal{P}}) = (1/6 + \delta{\cal{H}},
2\sqrt{2}+\delta{\cal{F}},
\sqrt{2} + \delta {\cal{P}} )$, and
eliminating ${\cal{H}}$ with Eq.(\ref{scalar_energy_exp}),
we find that the perturbation equations are
\beqn
\delta{\cal{F}}'  &=&  -{1\over 2}\delta{\cal{F}}+\delta{\cal{P}},
\label{calvprex}
\\
\delta{\cal{P}}'  &=&  -\delta{\cal{P}}+
\lambda \frac{\mu^4}{m_5^4} \exp\left({3\over
2}\tau-2\sqrt{2}\lambda e^{\tau/2} \right)
\nonumber \\&&-\frac{\sqrt{2}}{2}
\frac{\mu^4}{m_5^4} \exp
\left(\tau-2\sqrt{2}\lambda e^{\tau/2} \right).
\eeqn

In this case the time-dependent term in the equations drops very fast
as $\exp(-2\sqrt{2}\lambda \exp(\tau/2))$, we just analyze
the asymptotic equations
In the limit of $\tau\rightarrow \infty$, which is
\beqn
\delta{\cal{F}}'  &=&   -{1\over 2}\delta{\cal{F}}+\delta{\cal{P}},
\label{HFP_eq2}
\\
\delta{\cal{P}}'  &=&  -\delta{\cal{P}}
\label{HFP_eq3}
\eeqn
%
The eigenvalues of the system  (\ref{HFP_eq2}), (\ref{HFP_eq3}) are
-1/2 and -1, then the kinetic dominant solution is stable against
linear perturbations
for $V(\phi) = \mu^4 e^{-\lambda \phi/m_5}$ as well.
We can easily check that it is correct even including the time-dependent
terms.
\subsubsection{the inflationary solution}

Finally, we study a stability about the inflationary solution
(\ref{sevsr}) with the inverse-power-law potential
$V= \mu^{\alpha+4} \phi^{-\alpha} ~(\alpha<2)$.
We use $H, V(\phi)$ and $\dot{\phi}$ as our dynamical variables.
The asymptotic behavior of these variables for the inflationary solution
is
\beqn
{H\over m_5}&\sim& {V\over 6m_5^{4}}\sim {1\over
6 }{\cal V}_0
~(m_5t)^{-{\alpha\over 2}},
\\
{\dot{\phi}\over m_5^2}&\sim&{\cal P}_0 ~(m_5t)^{-{1\over 2}},
\eeqn
where ${\cal V}_0\equiv   (2\sqrt{\alpha})^{-\alpha}\left({\mu/
m_5}\right)^{\alpha+4} $ and  ${\cal P}_0 \equiv \alpha^{1\over 2}$.
For such time-dependence of the solution, we introduce  new
variables as
\beqn
{\cal{H}} &\equiv& m_5^{-1} H ~T^{{\alpha\over (2-\alpha)}},
\label{nnv1}\\
{\cal{V}} &\equiv& m_5^{-4}V ~T^{{\alpha\over (2-\alpha)}},
\label{nnv2}\\
{\cal{P}} &\equiv& m_5^{-2}\dot{\phi} ~T^{{1\over (2-\alpha)}},
\label{nnv3}\eeqn
where the new time coordinate $T$ is defined as
\beq
T \equiv (m_5 t)^{1-{\alpha\over 2}}.
\eeq
Using these variables, the basic equations are now,
\beqn
\left(1-{\alpha\over 2}\right)\frac{d {\cal{H}}}{d T}
&= &- {\cal{H}}
(6{\cal{H}} -{\cal{V}}) +\frac{\alpha}{2}{\cal{H}} ~T^{-1},
\\
\left(1-{\alpha\over 2}\right)\frac{d {\cal{V}}}{d T}
&=& \frac{\alpha}{2} {\cal{V}}
\left[ 1-\frac{\cal P}{{\cal P}_0}\left(\frac{\cal V}{{\cal
V}_0}\right)^{1/\alpha}
\right] T^{-1},
\\
\left(1-{\alpha\over 2}\right)\frac{d {\cal{P}}}{d T}
&=& - {{\cal P}{\cal V}\over 2}\left[\frac{6{\cal H}}{\cal{V}}
-\left(\frac{\cal P}{\cal P}_0\right)^{-1}\left(\frac{\cal V}{{\cal
V}_0}\right)^{1/\alpha}
\right]
\nonumber \\&&+\frac{\cal P}{2}~T^{-1},
\eeqn
with the following constraint equation,
\beq
{\cal H}=\frac{1}{12}{\cal P}^2 T^{-1} + \frac{1}{6}{\cal V}.
\label{scalar_energy_inf}
\eeq

In the limit of $T\rightarrow \infty$, we find the fixed point
$({\cal V}_0/6,{\cal V}_0, {\cal P}_0)$, which corresponds to the
inflationary solution.

Expanding
the basic equations around this fixed point as
$({\cal{H}}, {\cal{V}}, {\cal{P}})
= ({\cal{H}}_0 + \delta {\cal{H}},
{\cal{V}}_0 + \delta {\cal{V}},
{\cal{P}}_0 + \delta {\cal{P}})$, and eliminating
${\cal{H}}$ with Eq.(\ref{scalar_energy_inf}),
we find that the perturbation equations are
\beqn
\left(1-{\alpha\over 2}\right)\frac{d \delta {\cal V}}{d T}
&= &
-\frac{1}{2} \delta {\cal{V}} T^{-1}
-\frac{\alpha {\cal{V}}_0 }{2 {\cal{P}}_0 }\delta {\cal{P}} T^{-1},
\label{sc_in_eq1}
\\
\left(1-{\alpha\over 2}\right)\frac{d \delta {\cal P}}{d T}
&= &
\frac{{\cal P}_0}{2\alpha}\delta {\cal{V}}
-\frac{{\cal V}_0}{2}\delta {\cal{P}}\nonumber \\
&&+\frac{1}{2}{\cal P}_0(1-\frac{1}{2}{\cal P}_0^2) T^{-1}.
\label{sc_in_eq2}
\eeqn

If we adopt a naive analysis for the asymptotic equations ignoring
decaying time-dependent terms, we find that the eigen values are
$-{\cal V}_0/(2-\alpha)$ and 0.  The zero eigenvalue corresponds to
a marginally unstable mode, which appears in a slow-rolling phase of an
  inflationary solution.

In order to confirm the asymptotic behaviors of $\delta {\cal{V}} $ and
$\delta {\cal{P}}$, we will integrate the above equations directly.
To solve (\ref{sc_in_eq1}) and (\ref{sc_in_eq2}), we first need to find
the following  first-order differential equation
\beqn
\frac{d}{dT} {\large [} &&  A(T)~\delta {\cal{P}} +  B(T)~\delta
{\cal{V}}~{\large ]}=
\nonumber\\ &&\sigma \left[A(T)~\delta {\cal{P}} + B(T)~\delta
{\cal{V}}\right] + F(T).
\label{ap_eq0}
\eeqn
where $A(T)$, $B(T)$, and $F(T)$ are chosen appropriately as follows.

We can easily show that if $A(T)$, $B(T)$, $F(T)$ satisfy
\beqn
\frac{d A}{dT} &-& \left(\sigma + \frac{{\cal{V}}_0}
{2-\alpha}\right) A -
\frac{\alpha {\cal{V}}_0}{(2-\alpha) {\cal{P}}_0}
\frac{B}{T}=0,
\label{ap_eq1}
\\
\frac{d B}{dT} &+ & \frac{{\cal{P}}_0}{(2-\alpha)\alpha}
A-\left(\sigma+\frac{1}{(2-\alpha)T}
\right) \frac{B}{T}=0,
\label{ap_eq2}
\eeqn
and
\beqn
F = \frac{{\cal{P}}_0}{2-\alpha} \left( 1
-\frac{1}{2}{\cal{P}}_0^2\right)
{A\over  T} ,
\label{ap_eq_f}
\eeqn
(\ref{ap_eq0}) is found.

 From $(\ref{ap_eq1})$ and $(\ref{ap_eq2})$, by eliminating $A$,
we find the second order differential equation for
$B$, as,
\beqn
\frac{d^2 B}{d T^2} &&-
\left[\left(2\sigma +\frac{{\cal{V}}_0}{2-\alpha}\right)
+\frac{1}{2-\alpha}\frac{1}{T}\right]\frac{d B}{d T}+
\nonumber \\
&&
\left[\sigma\left(\sigma+ \frac{{\cal{V}}_0}{2-\alpha}\right)
+\left(\frac{\sigma}{2-\alpha}
+\frac{2 {\cal{V}}_0}{(2-\alpha)^2}\right)\frac{1}{T}\right.
\nonumber \\
&&\left.+\frac{1}{(2-\alpha)T^2}  \right]\times B = 0.
\label{ap_eq3}
\eeqn

The solution of this equation is given by two independent
hypergeometric functions, which  asymptotic behaviors are given by
\beqn
B_1(T) &=&e^{\sigma T} T^{\frac{2}{2-\alpha}},
\label{ap_eq_b1}
\\
B_2(T) &=&e^{[\sigma +\frac{{\cal{V}}_0}{2-\alpha}] T}
T^{-\frac{1}{2-\alpha}}.
\label{ap_eq_b2}
\eeqn

 From (\ref{ap_eq2}) and (\ref{ap_eq_f}),
we  find  $A(T)$ and $F(T)$ for each solution as
\beqn
A_1(T) &=& -\frac{\alpha}{{\cal{P}}_0} e^{\sigma T}
T^{\frac{\alpha}{2-\alpha}},
\label{ap_eq_a1}
\\
F_1 (T) &=& -{\alpha\over 2-\alpha}\left(1-\frac{1}{2}{\cal{P}}_0^2
\right) e^{\sigma T}  T^{2(\alpha-1)\over 2-\alpha},
\label{ap_eq_f1}
\\
A_2(T) &=& -\frac{\alpha {\cal{V}}_0}{{\cal{P}}_0}
e^{[\sigma +\frac{{\cal{V}}_0}{2-\alpha}] T}
T^{-\frac{1}{2-\alpha}},
\label{ap_eq_a2}
\\
F_2(T) &=& -{\alpha{\cal V}_0\over
2-\alpha}\left(1-\frac{1}{2}{\cal{P}}_0^2
\right)  e^{[\sigma +\frac{{\cal{V}}_0}{2-\alpha}]T}
T^{-\frac{3-\alpha}{2-\alpha}}.
\label{ap_eq_f2}
\eeqn

We then obtain two set of  (\ref{ap_eq0}), which general
solution
is given by
\beqn
A_1(T)~\delta {\cal{P}} + B_1(T)~\delta {\cal{V}}&=&e^{\sigma T}
\int^T dT' e^{-\sigma T'}~F_1(T'), \\
A_2(T)~\delta {\cal{P}} + B_2(T)~\delta {\cal{V}}&=&e^{\sigma T}
\int^T dT' e^{-\sigma T'}~F_2(T').
\eeqn

The asymptotic behaviors of these equations are
\beqn
-\frac{\alpha}{{\cal{P}}_0}
\delta {\cal{P}}&+&T \delta {\cal{V}}
\nonumber\\
&=&
c_1 T^{-\frac{\alpha}{2-\alpha}} - \left(1-\frac{1}{2}
{\cal{P}}_0^2\right)  ,
\label{ap_eq_al1}
\\
-\frac{\alpha {\cal{V}}_0}{{\cal{P}}_0}
  \delta {\cal{P}}
&+& \delta {\cal{V}}
\nonumber\\
&=&c_2 T^{\frac{1}{2-\alpha}} e^{-\frac{{\cal{V}}_0}{2-\alpha}T}
-\alpha \left(1-\frac{1}{2} {\cal{P}}_0^2\right)T^{-1} ,
\label{ap_eq_al2}
\eeqn
where $c_1$ and $c_2$ are integration constants.

Then  we   obtain the asymptotic solution of $\delta {\cal{P}}$ and
$\delta {\cal{V}}$ as
\beqn
\delta {\cal{P}} &=&
\Delta^{-1} {\large [}c_1
T^{-\frac{\alpha}{2-\alpha}}
-c_2  T^{\frac{3-\alpha}{2-\alpha}}e^{-\frac{{\cal{V}}_0}{2-\alpha}T}
\nonumber \\
&& +(\alpha-1)\left(1-\frac{1}{2} {\cal{P}}_0^2\right)
{\large ]},
\label{ap_eq_delp}
\\
\delta {\cal{V}} &=&
\Delta^{-1} {\large [} c_1 \frac{\alpha {\cal{V}}_0}
{{\cal{P}}_0}
T^{-\frac{\alpha}{2-\alpha}} -
c_2 \frac{\alpha}{{\cal{P}}_0}
T^{\frac{1}{2-\alpha}}e^{\frac{-{\cal{V}}_0}{2-\alpha}T}
\nonumber\\
&&-\frac{\alpha(\alpha-{\cal{V}}_0 T}{{\cal{P}}_0 T}
\left(1-\frac{1}{2} {\cal{P}}_0^2\right)
{\large ]},
\label{ap_eq_delv}
\eeqn
where
\beqn
\Delta = \frac{\alpha}{{\cal{P}}_0} \left[ {\cal{V}}_0T-1 \right].
\label{ap_eq_det}
\eeqn

(\ref{ap_eq_delp}), (\ref{ap_eq_delv})
and (\ref{ap_eq_det}) give  the
asymptotic behavior of $\delta {\cal{P}}$
and $\delta {\cal{V}}$ as
\beqn
\delta {\cal{P}} &\rightarrow&
{(\alpha-1){\cal{P}}_0\over \alpha {\cal V}_0}\left(1-\frac{1}{2}
{\cal{P}}_0^2\right)T^{-1} ,
\label{ap_eq6}
\\
\delta {\cal{V}} &\rightarrow&
-\left(1-\frac{1}{2} {\cal{P}}_0^2\right) T^{-1}
\label{ap_eq7}
\eeqn

This power-law decay corresponds to zero eigenvalue.
Then  the above naive analysis by eigen values is confirmed.
We conclude that  the inflationary solution is stable
against linear perturbations.

        \subsection{radiation dominated stage}

\subsubsection{the power-law solution}

We show here that for the scalar field with an inverse-power-law
potential model $V(\phi) = \mu^{4+\alpha} \phi^{-\alpha-4}$,
with $\alpha < 6$, the power-law solution in the
radiation dominant stage given by (\ref{sevkpc}), (\ref{phi0})
is stable against the linear perturbations.

Following \cite{Ratra_Peebles}, we introduce the new variables
as
\beqn
\tau &\equiv& \ln m_5 t\\
{\cal F} &\equiv& \frac{\phi}{\phi_e},\\
\label{def_u}
\eeqn
where $\phi_e$ is the exact solution given by (\ref{sevkpc}),
(\ref{phi0}),
\beqn
\frac{\phi_e}{m_5} = A(\alpha) (m_5 t)^{\frac{2}{\alpha+2}}.
\eeqn
$A(\alpha)$ is a dimensionless constant, which depends on
$\alpha$ and $\mu$,
\beqn
A(\alpha) = \left(\frac{2\alpha (\alpha +2)^2}{6-\alpha}\right)^
{\frac{1}{\alpha+2}} \left(\frac{\mu}{m_5}\right)^
{\frac{\alpha+4}{\alpha+2}}.
\eeqn

With these changes, Eq.(\ref{sderdeiv}) becomes,
\beqn
{\cal F}'&= &{\cal P} ,
\label{def_v}
\\
{\cal P}' &= &\frac{\alpha-14}{4(\alpha+2)}{\cal P}
+\frac{\alpha-6}{2(\alpha+2)^2}{\cal F}
-\frac{\alpha-6}{2(\alpha+2)^2}{\cal F}^{-\alpha-1},
\eeqn
where prime denotes the derivative with respect to $\tau$
and the power-law solution given by (\ref{sevkpc}) and (\ref{sevkpc})
corresponds to a fixed point $({\cal F},{\cal P})=(1,0)$.

In order to study the stability near the critical point,
we here use the linear analysis \cite{Ratra_Peebles}.
Expanding the above equations around the fixed point as
$({\cal F},{\cal P})=(1+\delta {\cal F}, \delta {\cal P})$, we find that
\beqn
\delta {\cal F}'&= &\delta {\cal P} ,
\\
\delta {\cal P}' &= & \frac{\alpha-6}{2(\alpha+2)}\delta {\cal F}
+ \frac{\alpha-14}{4(\alpha+2)} \delta {\cal P}.
\eeqn

Because these equations do not include the explicit
time-dependence, we can discuss the stability of the solution
by the eigenvalues of these coupled equations.

The eigenvalues are given as
\beqn
  - \frac{1}{2}
~~~ {\rm and} ~~~~-{6- \alpha \over \alpha+2}\eeqn
which are all negative for $\alpha<6$, and then this solution is stable
against linear perturbations.

\subsubsection{the kinetic dominant solution}

We show here that for the scalar field with an inverse-power-law
potential model $V(\phi) = \mu^{4+\alpha} \phi^{-\alpha-4}$,
with $\alpha > 6$, and an exponential potential model
$V(\phi) = \mu^4 e^{-\lambda \phi/m_5}$
the kinetic dominant solution in the
radiation dominant stage given by (\ref{rad_kin2}),
is stable against the linear perturbations.

At first, we show the case with an inverse-power-law
potential model $V(\phi) = \mu^{4+\alpha} \phi^{-\alpha-4}$,
with $\alpha > 6$.
As in the power-law solution in the radiation dominant stage,
we introduce the new variables
as
\beqn
\tau &\equiv& m_5 t,
\label{rad_kin_eq1}
\\
{\cal F} &\equiv& \frac{\phi}{\phi_k},
\label{rad_kin_eq2}
\eeqn
where $\phi_k$ is the kinetic dominant solution given by
(\ref{rad_kin2}),

\beqn
\frac{\phi_k}{m_5} = B (m_5 t)^{1/4}.
\label{rad_kin_eq3}
\eeqn

In this case,  $B$ is the dimensionless constant, whose value
depends on the initial condition.
In terms of these variables, the basic equation
(\ref{sderdeiv}) is expressed as,
\beqn
{\cal F}'&= &{\cal P},
\\
{\cal P}' &= & -\frac{1}{4} v +\alpha
\left(\frac{\mu}{m_5}\right)^{\alpha+4} B^{-\alpha-2} {\cal F}^{-\alpha-1}
e^{\frac{6-\alpha}{4}\tau}.
\eeqn

In the limit of $\tau \rightarrow \infty$, we find the critical point
$({\cal F}, {\cal P}) = ({\cal F}_0, 0)$, where ${\cal F}_0$ is an
arbitrary constant. This point corresponds to the kinetic dominant
solution. Expanding the basic equations around this critical point
as $({\cal F}, {\cal P}) = ({\cal F}_0 + \delta {\cal F}, \delta {\cal
P})$,  we find that the perturbation
equations are,
\beqn
\delta {\cal F}'&= & \delta {\cal P},
\\
\delta {\cal P}' &= &
-\frac{1}{4} \delta {\cal P} +\alpha
\left(\frac{\mu}{m_5}\right)^{\alpha+4} B^{-\alpha-2} {\cal
F}_0^{-\alpha-1} e^{\frac{6-\alpha}{4}\tau}.
\eeqn

Because these equations have explicit time-dependence,
we cannot evaluate the stability with the eigenvalues.
However, for $\alpha >6$,
we can easily integrate the above perturbed equations as

\beqn
\delta {\cal F} &=& \frac{16\alpha}{(7-\alpha) (6-\alpha)}
\left(\frac{\mu}{m_5}\right)^{\alpha+4} B^{-\alpha-2} {\cal
F}_0^{-\alpha-1} e^{\frac{(6-\alpha)\tau}{4}}
\nonumber\\
&-& 4\delta {\cal P}_0 e^{-\frac{\tau}{4}}+\delta {\cal F}_0,
\\
\delta {\cal P} &=& \frac{4\alpha}{7-\alpha}
\left(\frac{\mu}{m_5}\right)^{\alpha+4} B^{-\alpha-2} {\cal
F}_0^{-\alpha-1} e^{\frac{(6-\alpha)\tau}{4}}
\nonumber\\
&+& \delta {\cal P}_0 e^{-\frac{\tau}{4}},
\eeqn
for $\alpha \neq 7$, and

\beqn
\delta {\cal F} &=& -28\left(\frac{\mu}{m_5}\right)^{11} B^{-9} {\cal
F}_0^{-8}  e^{-\frac{\tau}{4}} (\tau - 4)
\nonumber\\
&-& 4\delta {\cal P}_0 e^{-\frac{\tau}{4}} +\delta {\cal F}_0,
\\
\delta {\cal P} &=& 7\left(\frac{\mu}{m_5}\right)^{11} B^{-9} {\cal
F}_0^{-8}
\tau e^{-\frac{\tau}{4}} + \delta {\cal P}_0 e^{-\frac{\tau}{4}},
\eeqn
for $\alpha = 7$,
where $\delta {\cal F}_0$ and $\delta {\cal P}_0$ are integration
constants.

These show that in the limit of  $\tau \rightarrow \infty$,
even though this is not strictly a fixed point,
the kinetic-term dominant
solutions are attractors in a sense that these satisfy
(\ref{rad_kin2}).

Next,  we investigate the case with an exponential
potential model $V(\phi) = \mu^4 e^{-\lambda \phi/m_5}$.

In this case, the equation of motion for the scalar field
(\ref{sde}) is now,
\beqn
\ddot{\phi} + \frac{3}{4t} \dot{\phi} -\lambda \mu^3
\frac{\mu}{m_5} \exp [-\lambda \phi/m_5]=0.
\label{rad_kin_eq4}
\eeqn

In terms of the variables, (\ref{rad_kin_eq1})-
(\ref{rad_kin_eq3}), the basic equation
(\ref{rad_kin_eq4}) is expressed as,
\beqn
{\cal F}'&= &{\cal P},
\\
{\cal P}' &= & -\frac{1}{4} {\cal P}
+\frac{\lambda}{B} \left(\frac{\mu}{m_5}\right)^4\exp
\left[\frac{7}{4} \tau - \lambda B e^{\frac{1}{4}\tau} {\cal F}\right].
\eeqn

In the limit of $\tau \rightarrow \infty$, we find the critical point
$({\cal F}, {\cal P}) = ({\cal F}_0, 0)$, where ${\cal F}_0$ is an
arbitrary constant. This point corresponds to the kinetic dominant
solution. Expanding the basic equations around this critical point
as $({\cal F}, {\cal P}) = ({\cal F}_0 + \delta {\cal F}, \delta {\cal
P})$,  we find that the perturbation
equations are,
\beqn
\delta {\cal F}'&= & \delta {\cal P},
\\
\delta {\cal P}' &\approx &
-\frac{1}{4} \delta {\cal P}
+\frac{\lambda}{B} \left(\frac{\mu}{m_5}\right)^4\exp
\left[\frac{7}{4} \tau - \lambda B e^{\frac{1}{4}\tau} {\cal F}_0\right].
\eeqn

In the limit of $\tau \rightarrow \infty$,
the asymptotic behavior of
these equations are,
\beqn
\delta {\cal F}'&= & \delta {\cal P},
\\
\delta {\cal P}' &\approx &-\frac{1}{4} \delta {\cal P},
\eeqn
which can be integrated as

\beqn
\delta {\cal F}&= & \delta {\cal F}_0 -4 \delta {\cal P}_0 e^{-\tau/4},
\\
\delta {\cal P} &\approx &\delta {\cal P}_0 e^{-\tau/4},
\eeqn
where $\delta {\cal F}_0$ and $\delta {\cal P}_0$ are integration
constants.

These show that in the limit of  $\tau \rightarrow \infty$,
even though this is not strictly a fixed point,
the kinetic term dominant
solutions are attractors in a sense that these satisfy
(\ref{rad_kin2}).

        \section{Global stability of the solutions}
\label{sec.5}

In the previous section, we show that the analytic solutions
listed up in Table 1, are stable against the linear perturbations.
However, in order to show that these solutions
are realized from wide initial conditions,
we have to study stability against non-linear perturbations.
The we analyze a global stability of the solutions
in the phase-space.

For the model with the potential $V(\phi)= \mu^6 \phi^{-2}$, we study it
  both numerically and analytically
including both the scalar field and the radiation (\S\ref{5-A}).
We also show the numerical results
for the kinetic dominant solutions (\ref{sevkd})
and the inflationary solutions (\ref{sevsr})
by assuming  scalar field dominance (\S\ref{5-B}) and
for the power-law solutions ($\alpha \neq 2$) (\ref{sevkpc})
and the kinetic dominant solutions (\ref{rad_kin2})
by assuming  radiation dominance (\S\ref{5-C}).

        \subsection{the power-law solution ($\alpha = 2$)
\label{5-A}}

Here, we study a global stability of the solutions  for
$V(\phi) = \mu^6 \phi^{-2}$ in detail, following \cite{cope_et.al}.
We introduce new variables
\beqn
X &=& \frac{m_5^{-3/2}}{2\sqrt{3}} \frac{\dot{\phi}}{\sqrt{H}},
\nonumber \\
Y &=& {2m_5^{3/2} \over \sqrt{3}} {\cal M}_0\frac{\phi^{-1}}{\sqrt{H}},
\eeqn
where ${\cal M}_0 \equiv \mu^3/(2\sqrt{2}m_5^3)$,
and new derivative with respect to $\ln a$, which is described by  $'$.
The dynamical equations are rewritten as a plane-autonomous system:
\beqn
X' =&F(X,Y)\equiv & X (X^2 - 2 Y^2 - 1 ) + 3{\cal M}_0^{~-1}Y^3,\nonumber \\
Y' =&G(X,Y)\equiv & -Y\left( 2Y^2 - X^2 +3{\cal M}_0^{~-1}XY -2\right),
\label{eq:xy}
\eeqn
and radiation energy is given by the constraint equation
\beqn
\frac{\rho_r}{6 m_5 ^3 H}+ X^2 + Y^2 = 1.
\eeqn

 From the constraint equation, the density parameter of a scalar field is 
given by
\beqn
\Omega_\phi \equiv \frac{\rho_\phi}{6 m_5 ^3 H} = X^2 + Y^2.
\eeqn

Since radiation energy is non-negative ($\rho_r \geq 0$),
we find  $0 \leq X^2 + Y^2 \leq 1$,
so the evolution of this system is completely described
by trajectories within the unit disc.
As the system is symmetric under the reflection $(X,Y) \rightarrow  (X,-Y)$ and
the evolution will not go beyond the $Y=0$-line (which corresponds to
$\phi=\infty$ or $H=\infty$),
it is enough to discuss
the upper half-disc ($Y \geq 0$).

Depending on  the value of ${\cal M}_0 ({\rm or} \mu / m_5)$, we have four or
five fixed points (critical points) where $X'$ and $Y'$ vanish.
The four critical points $(0,0),$ $(1,0),$ $ (-1,0)$, and
A$ (1/ \sqrt{1+{\cal M}_0^{~2}},$ $ {\cal M}_0/ \sqrt{1+{\cal
M}_0^{~2}})$  exist on the boundary of the phase space, while the fifth
critical point B
$ (\sqrt[4]{8/9}{\cal M}_0^{1/2},$ $ \sqrt[4]{2/9}{\cal
M}_0^{1/2})$ appears in the half unit disc only if ${\cal M}_0<\sqrt{2}/2$.

The simple analysis by linear perturbations shows that the critical point
A is stable if ${\cal M}_0>\sqrt{2}/2$, while it becomes a saddle point
when ${\cal M}_0<\sqrt{2}/2$, and newly appeared critical point B becomes
stable. Since the point A locates on the boundary, $\Omega_\phi$ is
always unity (scalar-field dominated stage), while the point B appears on
the line $Y=X/\sqrt{2}$. The density parameter of this solution is
$\Omega_\phi = \sqrt{2}{\cal M}_0$, which means that the radiation
density becomes larger that that of a scalar field if  ${\cal
M}_0<\sqrt{2}/4$

\begin{figure}[htbp]
\epsfysize =8.0cm
\epsfbox{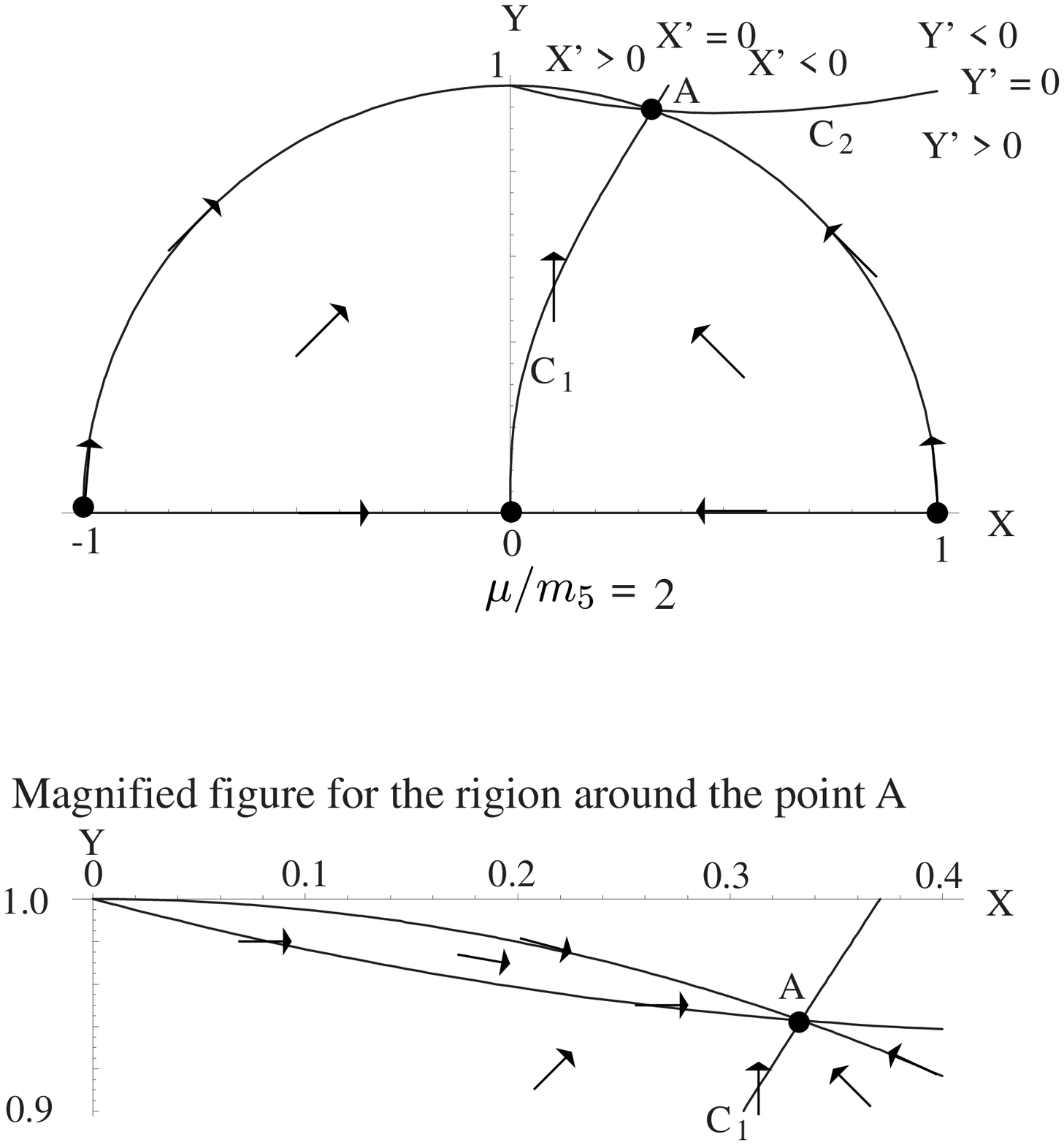}
\vskip 1em
\caption{Schematic phase diagram for the
model with $V(\phi) =\mu^6
\phi^{-2}$ for the case of  ${\cal M}_0>\sqrt{2}/2$.
The arrows show the directions in which the spacetime evolves.
  The critical point A  is an attractor.
}
\label{Figpowerlawphase}
\end{figure}

\begin{figure}[htbp]
  \leavevmode
\epsfysize = 8.0cm
\epsfbox{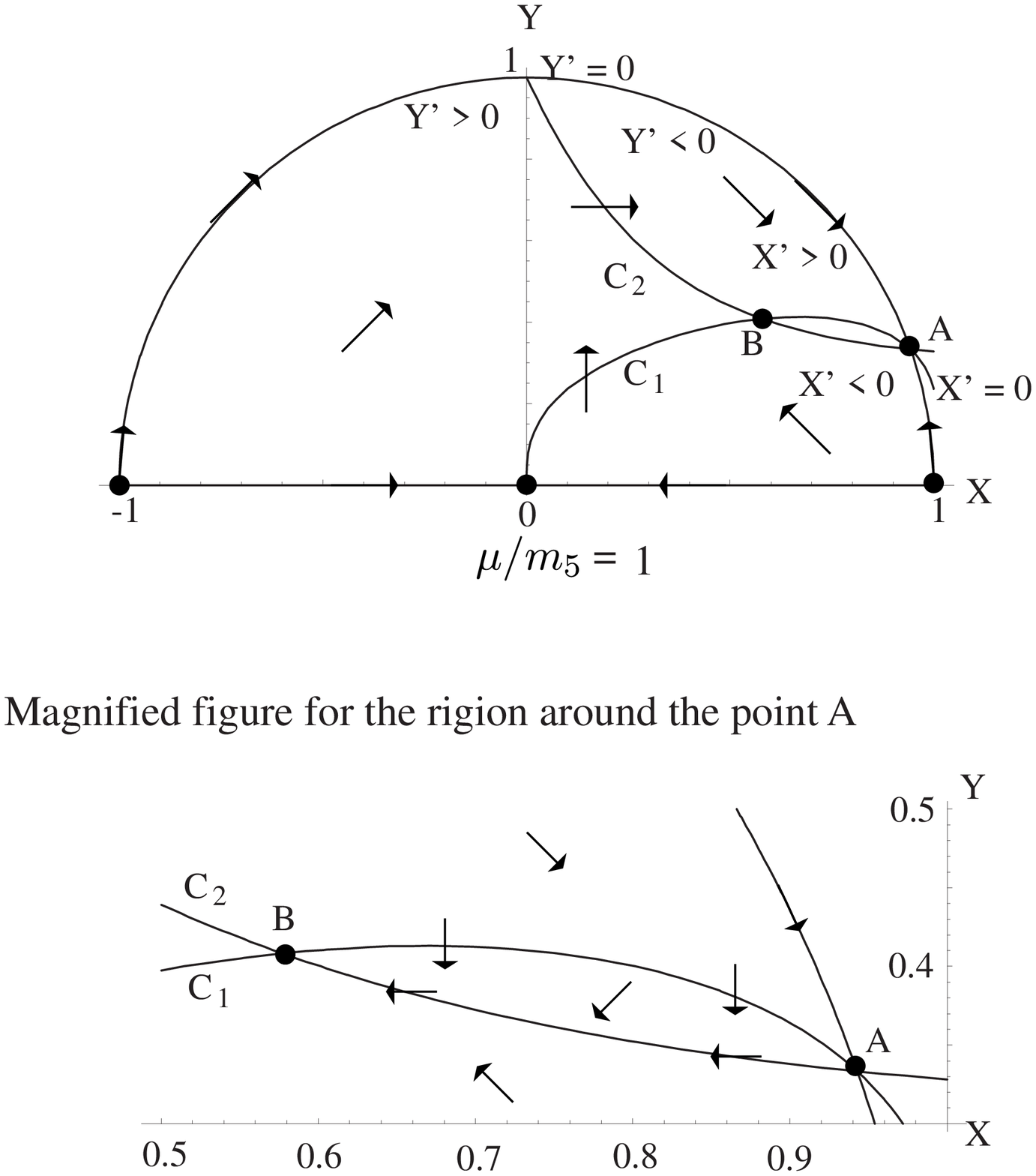}
\vskip 1em
\caption{Schematic phase diagram for the
model with $V(\phi) =\mu^6
\phi^{-2}$ for the case of  ${\cal M}_0<\sqrt{2}/2$.
The arrows show the directions in which the spacetime evolves.
  The critical point A  is a saddle point, while the point B is an
attractor. }
\label{Figscalingphase}
\end{figure}

In order to analyze the global behavior in the phase space of the present
dynamical system, we first draw critical  curves corresponding to
$F(X.Y)=0$, which corresponds to  the curve $C_1$ or
$G(X,Y)=0$, which  gives two curves;the straight
line $Y=0$ and hyperbolic curve $C_2$ (see Fig. \ref{Figpowerlawphase} and
Fig.\ref{Figscalingphase}). 

From the signs
of $F(X.Y)$ and $G(X,Y)$, We find the direction of the evolutionary track
of the universe in the phase space,
which is shown by  arrows in Fig. \ref{Figpowerlawphase} and
Fig.\ref{Figscalingphase}.  Following those arrows, we find
whether the critical points are globally stable or unstable.
It turns out that the critical points
$(1,0)$  and
$ (-1,0)$  are unstable, and $(0,0)$ is a saddle point.
For ${\cal M}_0>\sqrt{2}/2$, the
critical point A is stable, but  if ${\cal M}_0<\sqrt{2}/2$, it becomes a
saddle point, as we expect from our perturbation analysis, and new
critical point B becomes a stable point.

We also show the result of numerical analysis for both cases.
We set $\mu = 2m_5 ({\cal M}_0=2\sqrt{2})$
for the case of ${\cal M}_0>\sqrt{2}/2$ and  $\mu = m_5  ({\cal
M}_0=\sqrt{2}/4)$ for the case of ${\cal M}_0<\sqrt{2}/2$.

\begin{figure}[htbp]
  \leavevmode
\epsfysize =4.5cm
\epsfbox{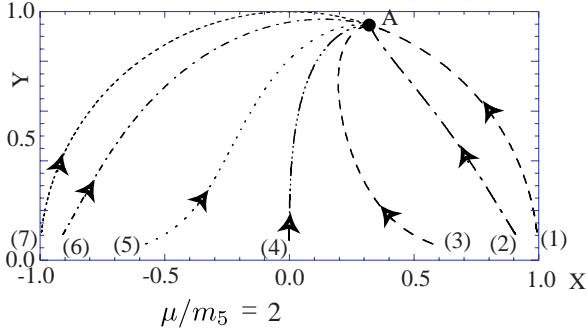}
\vskip 1em
\caption{Numerical analysis for the
model
$V(\phi) =
\mu^6
\phi^{-2}$ with $\mu = 2m_5$ $ ({\cal M}_0=2\sqrt{2}>\sqrt{2}/2)$.
  The critical point
$(X, Y) = (1/3, 2\sqrt{2}/3)$
corresponding to the power-law solution (\ref{plif})
and (\ref{sevpl}) is the attractor.
As for initial values of ($\phi/m_5$,
$(\phi/m_5)' \equiv t \times d( \phi/m_5)/dt, \rho_r/m_5^4)$ at  $m_5 t =
1.0$, we set ~
(1) (1.0 $\times 10^{2}$, 1.0, 1.0 $\times 10^{-4})$,~
(2) (1.0 $\times 10^{2}$, 1.0, 1.0 $\times 10^{-1})$,~
(3) (1.0 $\times 10^{2}$, 1.0, 1.0 ),~
(4) (1.0 $\times 10^{2}$, 0.0, 1.0 ),~
(5) (1.0 $\times 10^{2}$, -1.0, 1.0),~
(6) (1.0 $\times 10^{2}$, -1.0 , 1.0 $\times 10^{-1})$,~
(7) (1.0 $\times 10^{2}$, -1.0 , 1.0 $\times 10^{-4})$.
}
\label{Figpowerlaw}
\end{figure}

\begin{figure}
  \epsfysize=4.5cm
\epsfbox{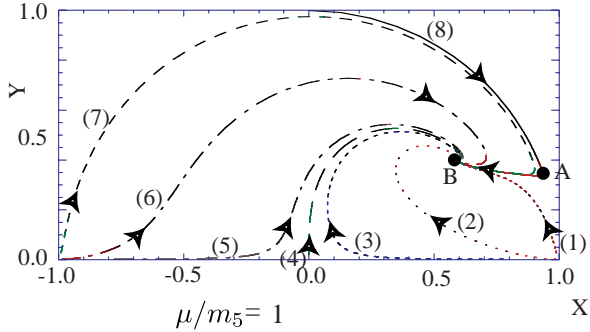}
\vskip 1em
\caption{
Numerical analysis for the
model  $V(\phi) = \mu^6
\phi^{-2}$ with $\mu = m_5$ $ ({\cal M}_0=\sqrt{2}/4<\sqrt{2}/2)$.  The
critical point
$(X, Y) = (2\sqrt{2}/3, 1/3)$ corresponding to the power-law solution
(\ref{sevkpc}) is the attractor. As for initial values of ($\phi/m_5$,
$(\phi/m_5)' \equiv t \times d( \phi/m_5)/dt, \rho_r/m_5^4)$ at  $m_5 t =
1.0$, we set ~
(1) (5.0 $\times 10^{1}$, 1.0 , 1.0 $\times 10^{-2})$,~
(2) (5.0 $\times 10^{3}$, 1.0 , 1.0 $\times 10^{-2})$,~
(3) (5.0 $\times 10^{5}$, 1.0 , 1.0 $\times 10^{-2})$,~
(4) (5.0 $\times 10^{1}$, 0.0, 1.0 ),~
(5) (5.0 $\times 10^{5}$, -1.0 , 1.0 $\times 10^{-2})$,~
(6) (5.0 $\times 10^{3}$, -1.0 , 1.0 $\times 10^{-2})$,~
(7) (5.0 $\times 10^{1}$, -1.0 , 1.0 $\times 10^{-2})$,~
(8) (5.0 $\times 10^{-1}$, 0.0, 1.0 $\times 10^{-2})$.
}
\label{Figscaling}
\end{figure}

In summary, we find that for $(\mu/m_5)^6 > 4$,
the power law solution corresponding to
the critical point A is an attractor, while for
$(\mu/m_5)^6 < 4$ the scaling solution corresponding to the critical
point B becomes an attractor.

        \subsection{scalar field dominated stage\label{5-B}}

In what follows, we only show numerical analysis for global stability for
other cases.

        \subsubsection{the kinetic dominant solution}

\begin{figure}
\epsfysize =6.0cm
\epsfbox{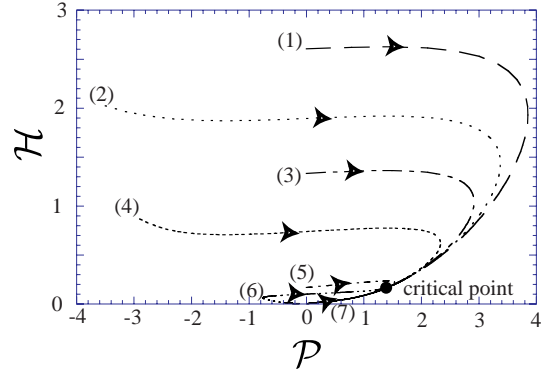}
\caption{
Global stability of the solutions for the model with $V(\phi) = \mu^7
\phi^{-3}$.
This shows that the kinetic dominance
(on the hyperbola ${\cal{H}} = (1/12){\cal{P}}^2$)
realizes at first, then the solution eventually approaches the critical  point
( ${\cal{H}}=1/6$, ${\cal{P}}= \sqrt{2}$), which corresponds to
the solution (\ref{sevkd}).
We set $\mu= m_5$.
As for initial conditions ($\phi/m_5$,
$(\phi/m_5)' \equiv t \times d( \phi/m_5)/dt)$ at $m_5 t =1.0$, we
choose the following seven sets of initial data;~
(1) (4.0 $\times 10^{-1}$, 0.0), (2) (5.5 $\times 10^{-1}$, -3.5),~
(3) (5.0 $\times 10^{-1}$, 0.0), (4) (1.0, -2.9),~
(5) (1.0 , 0.0), (6) (2.0 $\times 10^{1}$, 1.0$\times 10^{-1}$),~
(7) (1.0 $\times 10^{1}$, 0.0).
}
\label{Figkinetic}
\end{figure}

First, we show that the kinetic dominant solution  (\ref{sevkd})
is obtained as an attractor  from the various
initial conditions  (see Fig.\ref{Figkinetic}). The  kinetic dominant solution
is described by  ${\cal{H}} = (1/12){\cal{P}}^2$, and the solution of
(\ref{sevkd}) is given by
${\cal{H}}=1/6$, ${\cal{P}}= \sqrt{2}$. 
(${\cal{H}}$ and ${\cal{P}}$ are defined by 
Eq.(\ref{nv1}) and Eq.(\ref{nv3}), respectively.)
Even though we show only the
case for the model $V(\phi) = \mu^7 \phi^{-3}$, the similar results  are
obtained for the model with $V(\phi) = \mu^{4+\alpha} \phi^{-\alpha}$,
$\alpha > 2$, and $V(\phi) = \mu^4 \exp[-\lambda \frac{\phi}{m_5}]$.
Fig. \ref{Figkinetic} shows that the kinetic dominance 
realizes at first, then the solution eventually approaches the critical point
(${\cal{H}}=1/6$, ${\cal{P}}= \sqrt{2}$),

        \subsubsection{the inflationary solution}

\begin{figure}
\epsfysize =6.0cm
\epsfbox{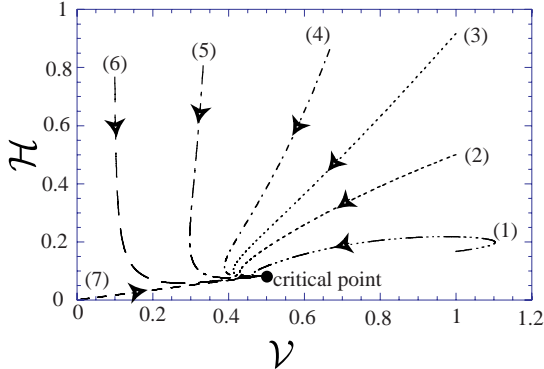}
\caption{
Global stability of the solutions for the model with $V(\phi) = \mu^5
\phi^{-1}$.
This shows that the slow-rolling (${\cal{H}} = (1/6)
{\cal{V}}$) realizes  first, then the solution eventually
approaches the critical point  (${\cal{V}} = 1/2$,
${\cal{H}}  = 1/12$), which corresponds to the solution
(\ref{sevsr}). We set $\mu= m_5$. As for initial conditions
($\phi/m_5$,
$(\phi/m_5)' \equiv t \times d( \phi/m_5)/dt)$ at $m_5 t = 1.0$, we
choose the following seven sets of initial data;\\
(1) (1.0, 0.0), (2) (1.0, 2.0),
(3) (1.0, 3.0),\\ (4) (1.5, 3.0),
(5) (3.0, 3.0), (6) (1.0$\times 10^1$, 3.0 ),
(7) (1.0 $\times 10^2$, 0.0).
}
\label{Figinflation}
\end{figure}

Next, we show the inflationary solution  (\ref{sevsr})
realizes from the various initial
conditions  numerically (see Fig.\ref{Figinflation}).
The slow-rolling condition gives 
${\cal{H}} = (1/6) {\cal{V}}$, and the solution
(\ref{sevsr}) is described by
${\cal{H}}  = 1/12$,
${\cal{V}}  = 1/2$.
(${\cal{H}}$ and ${\cal{V}}$ are defined by 
Eq.(\ref{nnv1}) and Eq.(\ref{nnv2}), respectively.)
From Fig. \ref{Figinflation}, we easily find the inflationary solution for a
wide range of initial data as expected.
 Even though
we show only the case for the model with $V(\phi) = \mu^5 \phi^{-1}$,  the
similar results  are obtained for the model with $V(\phi) = \mu^{4+\alpha}
\phi^{-\alpha}$, $\alpha < 2$.

        \subsection{radiation dominated stage\label{5-C}}

        \subsubsection{the power-law solution $(\alpha \neq 2)$}

\begin{figure}
\epsfysize =6.0cm
\epsfbox{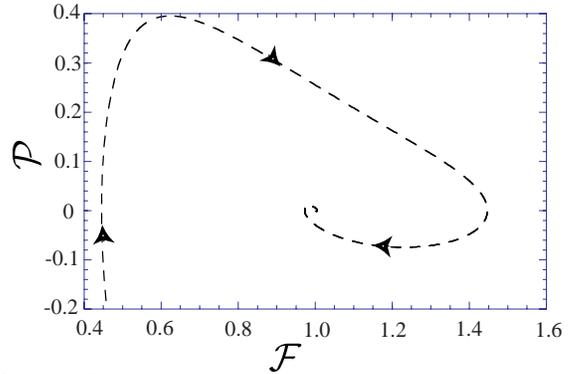}
\caption{
The attractor behavior of the power-law solution.
We set $\mu= m_5$.
As for initial condition, we start
from $m_5 t = 1.0$. The initial values of $({\cal F},{\cal P})$ are \\
$(4.6\times 10^{-1}, -1.8\times 10^{-1})$, 
which corresponds to the initial values of
($\phi/m_5$, $(\phi/m_5)' \equiv t \times d( \phi/m_5)/dt)$ are
(1.0, 0.0).
Furthermore, $\rho_\phi / m_5^4 = 1.0$
and $\rho_r / m_5^4 = 1.0$.
}
\label{Figradpowerlaw}
\end{figure}

Here, we show that the power-law solution
(\ref{sevkpc}) and (\ref{phi0})
($\alpha \neq 2$)
realizes from the
distant point in the parameter space numerically
(see Fig.\ref{Figradpowerlaw}). The power-law
solution is given by  $({\cal F},{\cal P}) = (1,0)$. Because this system can be
written as a plane-autonomous system, and there are no other stable fixed
point, we only show one example. Even though we show only the case for the
model with 
$V(\phi) = \mu^7 \phi^{-3}$, the similar results are obtained for the model
with $V(\phi) = \mu^{4+\alpha} \phi^{-\alpha}$, $2 < \alpha < 6$.

        \subsubsection{the kinetic dominant solution}

\begin{figure}
\epsfysize =6.0cm
\epsfbox{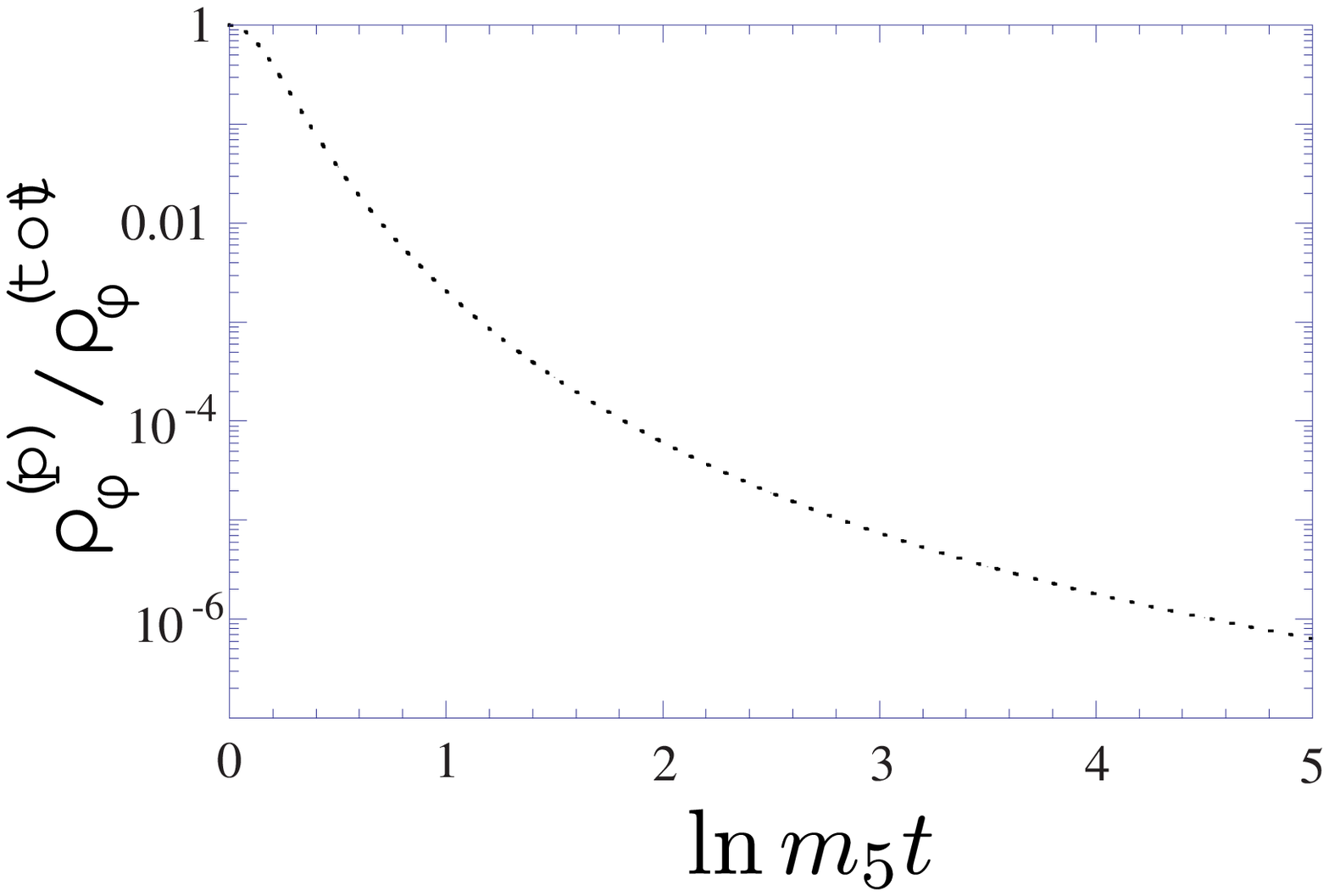}
\caption{
The
ratio of the potential energy of the scalar field to
the total energy of the scalar field.
This shows that even though the potential energy dominates the kinetic
energy initially, the kinetic term dominance eventually realizes.
We set
$\mu= m_5$.
As for initial condition, we start
from $m_5 t = 1.0$. The initial values of
($\phi/m_5$, $(\phi/m_5)' \equiv t \times d( \phi/m_5)/dt)$ are
$(1.2, 0.0)$.
Furthermore, $\rho^{(tot)}_\phi / m_5^4 = 2.8 \times 10^{-1}$
and $\rho_r / m_5^4 = 1.0$
}
\label{Figradkin}
\end{figure}

Finally, we show the result  for the models with 
$V(\phi) = \mu ^{4+\alpha} \phi^{-\alpha}$, with $\alpha > 6$ and
$V(\phi) = \mu^4 e^{-\lambda \phi/m_5}$, in the radiation dominant
stage (Fig.\ref{Figradkin}).
Because this kinetic dominant solution does not include
the characteristic mass scale,
we cannot introduce the fixed critical point like the previous solutions.
Therefore, we show the
ratio of the potential energy of the scalar field to
the total energy of the scalar field.
 From the potential dominant initial condition
($\rho^{(p)}_\phi/\rho^{(tot)}_\phi=1$ initially),
the kinetic term eventually dominates the potential term.
($\rho^{(p)}_\phi/\rho^{(tot)}_\phi \rightarrow 0$).
Even though we show only the case for
the model $V(\phi) = \mu^{11} \phi^{-7}$, the similar results are obtained for
the other models
$V(\phi) = \mu ^{4+\alpha} \phi^{-\alpha}$, with $\alpha > 6$ and
$V(\phi) = \mu^4 e^{-\lambda \phi/m_5}$.

\section{Summary and Discussion}
\label{sec.6}

In this paper we have studied the dynamics of a scalar field
with the typical potential in a brane world scenario.
We have adopted the second Randall-Sundrum brane scenario and assumed
that the scalar field is confined on our 4-dimensional spacetime. In this
model, because of the quadratic term of the energy density, which is from the
consequence  of our brane embedded in the extra dimension, the dynamics of the
scalar field is modified  from a conventional cosmology in the early stage of
the universe. 

For the power-law potential model,
which was discussed in \cite{maart_et.al},
there exist an inflationary solution
in the quadratic term dominant stage, which satisfies  slow-rolling
condition. Similar to the  conventional cosmology, the $e$-folding number
more than $60$ is obtained for sufficiently large initial values of
$\phi$. However, because the cosmic expansion  is faster
than  that in the conventional cosmology, a shorter time is required
for the enough
$e$-folding number.

After the slow-rolling condition is broken, and the oscillating phase
starts, the dynamics of the scalar field depends on the power of the scalar
field
potential.
For the model, $V(\phi) = \frac{1}{2} m^2 \phi^2$, the energy density  of
the scalar field  decreases as the pressureless perfect fluid and more
slowly than that in the conventional case. In the preheating stage, this
changes the time dependence of the variable appeared  in the Mathieu
equation and leads to an efficient particle production via non-perturbative
decay of inflaton even  for the smaller coupling constant.
For the model, $V(\phi) = \frac{1}{4}\lambda \phi^4$, the energy density
of the scalar field decreases as the radiation fluid. In this case, the
solution of the scalar field evolution is represented in terms of the
elliptic function and the form of the Lam$\acute{e}$ equation in the
preheating stage is not changed. Although  this is the same as
the conventional case, because of the slower growth of the scale factor,
the particle production becomes a little efficient\cite{tsujikawa}.

For the models with an inverse-power-law potential model ($V = \mu^{4+\alpha}
\phi^{-\alpha}$) or an exponential potential model
$V = \mu^4 e^{-\lambda \phi/m_5}$, we find much difference from 
conventional cosmology.
We find analytic solutions in several circumstances, and analyze those
stabilities.
According to the solutions and those analysis, the aymptotic behavior of the
universe is sammarized as follows:

For  $V = \mu^{4+\alpha} \phi^{-\alpha}$, with  $\alpha < 2$ and
with $\alpha=2$ ($\mu> \sqrt[6]{40}m_5$), an inflationary solution
is obtained  in the quadratic term dominant stage.
Since this is an attractor, such a solution is eventually obtained  even
though from the radiation-dominant initial condition.
The inflationary solutions   in these models are milder than that
  in the conventional cosmology. Because this potential has no
minima, in order to recover the  Big-Bang cosmology,  some
unknown reheating mechanisms is required . The gravitational particle
production could provide its resolution\cite{g.p.prod.}.

For  $V = \mu^6 \phi^{-2}$ with $\sqrt[6]{4}~ m_5< \mu<\sqrt[6]{40}~ m_5$,
 we find one critical  solution, i.e. 
$a \sim t^p$, with $p=[1+1/8(\mu/m_5)^6]/6$.
The power $p$ is in the range of 
$1/4 < p < 1$. This solution is the attractor for
$\mu > \sqrt[6]{4}~m_5$,
which gives
$\Omega_\phi=1$ and
$1/4<p<1$. While  for $\mu<\sqrt[6]{4}~ m_5$, 
another critical point with $p=1/4$ appears, which turns to be the attractor.
For this scaling solution, $\Omega_\phi$ depends on
$\mu$ 
($\Omega_\phi= \frac{1}{2}(\frac{\mu}{m_5})^3$), 
and  the small value of $\mu$ ($ \mu <\sqrt[6]{4} ~m_5$),
the energy density of radiation in this attractor universe exceeds that
of the scalar field. 

For $V = \mu^{4+\alpha} \phi^{-\alpha}$,
with  $2 < \alpha <6$, which we
 discussed in the previous paper for the quintessence scenario
\cite{maeda,mizuno_maeda}, the attractor
universe is  radiation dominant and the power-law solution is the
attractor for the scalar field. Our work concerned to quintessence
for this model is completed in this paper to show the stability of the
solutions.  (For the naturalness of the quintessence scenario
and the constraints to this model, see \cite{maeda,mizuno_maeda}.).

For $V = \mu^{4+\alpha} \phi^{-\alpha}$, with  $6 < \alpha$ and
$V= e^{-\lambda \phi/m_5}$, because the shape of the potential is too
steep, the kinetic term of the scalar field dominates the potential term
both in the scalar field dominant universe and the radiation dominant universe
asymptotically.

In this paper, we have assumed that a scalar field is confined in the
brane. However, most models based on a string theory
require a dilaton field in the bulk, which may gives
very important effects on the behavior of gravity and a scalar field on the
brane\cite{maeda_wands}.
The analysis in such models is left to the
future work.

\section*{ACKNOWLEDGMENTS}
We would like to thank T. Harada, and S. Tsujikawa for useful
discussions and comments.  This work was  partially supported by the 
Grant-in-Aid for Scientific Research Fund of the Ministry of Education,
Science and Culture (Nos. 14047216, 14540281) and by the Waseda University
Grant for Special Research Projects and  by the Yamada foundation.


\begin{thebibliography}{99}

\bibitem{inflation}
A. Linde,
{\em Particle Physics and Inflationary  Cosmology},
(Harwood academic publishers, 1980);
A. R. Liddle, and D. H. Lyth,
{\em Cosmological  Inflation  and Large-Scale  Structure},
(Cambridge University Press, Cambridge, 2000).

\bibitem{defect}
A. Vilenkin, Phys. Rep.{\bf 121}, 263 (1985);
T. W. B. Kibble, J. Phys. {\bf A9},1387 (1976);
J. Preskill, Ann. Rev. Nucl. Part. Sci. {\bf 34},461(1984).

\bibitem{Ratra_Peebles}
B. Ratra and P. J. E. Peebles, Phys. Rev. D {\bf 37}, 3406 (1988).

\bibitem{Quintessence}
R. R. Caldwell, R. Dave and P. J. Steinhardt, Phys. Rev. Lett. {\bf 80},
1582 (1998);I. Zlatev, L. Wang and P. J. Steinhardt, Phys. Rev. Lett.
{\bf 82}, 896 (1999); P. J. Steinhardt, L. Wang and I. Zlatev, Phys. Rev.
D {\bf 59}, 123504 (1999).

\bibitem{SUGRA}
B. Whitt, Phys. Lett. B {\bf145}, 176 (1984);
J. D. Barrow and S. Cotsakis,  Phys. Lett. B {\bf214}, 515 (1988);
D. Wands, Class. Quantum Grav. {\bf11},269 (1994);
M. B. Green, J. H. Schwarz, and E. Witten, {\it Superstring Theory}
( Cambridge University Press, Cambridge, England, 1987).

\bibitem{3-brane}
V. A. Rubakov and M. E. Shaposhinikov, Phys. Lett. B {\bf 125}, 139
(1983); K. Akama, in  {\it Gauge Theory and Gravitation} ed by K.
Kikkawa, N. Nakanishi, and H. Nariai  (Springer-Verlag, 1983); K. Akama,
hep-th/0001113.
\bibitem{String}
N. Arkani-Hamed, S. Dimopoulos and G. Dvali,  Phys. Lett. B {\bf 429},
263 (1998);  I. Antoniadis, N. Arkani-Hamed, S. Dimopoulos and G. Dvali,
Phys. Lett. B {\bf 436}, 257 (1998).
\bibitem{H-W}
P. Ho$\check{\rm r}$ava and E. Witten, Nucl. Phys. B {\bf 460},  506
(1996); ibid B {\bf 475}, 94 (1996).
\bibitem{RS}
L. Randall and R. Sundrum, Phys. Rev. Lett. {\bf 83},  4690 (1999);
L. Randall and R. Sundrum, Phys. Rev. Lett. {\bf 83},  3370 (1999).
\bibitem{SMS}
T.  Shiromizu, K.  Maeda, and M.  Sasaki, Phys.  Rev.  D {\bf 62},
024012 (2000).
\bibitem{Brev}
R. Maartens, gr-qc/0101059.
\bibitem{Bcmgy1}
P. Bin$\acute{e}$truy, C. Deffayet and D. Langlois, Nucl. Phys. B  {\bf
565}, 269 (2000); N. Kaloper, Phys.  Rev.  D {\bf 60}, 123506 (1999) ; C.
Csaki, M. Graesser, C. Kolda  and J. Terning Phys. Lett. B {\bf 462}, 34
(1999) ; T. Nihei, Phys. Lett. B {\bf 465},  81 (1999) ; P. Kanti, I. I.
Kogan, K. A. Olive and M. Prospelov, Phys. Lett. B {\bf 468}, 31 (1999);
J. M. Cline, C. Grojean and G. Servant, Phys. Rev. Lett. {\bf 83},  4245
(1999);  P. Bin$\acute{e}$truy, C. Deffayet, U. Ellwanger and D. Langlois,
Phys. Lett. B {\bf 477}, 285 (2000) ;  S. Mukohyama, T. Shiromizu and K.
Maeda, Phys. Rev. D {\bf 62}, 024028 (2000).
\bibitem{Bcmgy2}
      L. Mendes and A. R. Liddle, Phys. Rev. D {\bf 62},
103511 (2000);  J. Khoury, P. J. Steinhardt and D. Waldram, Phys.  Rev. D
{\bf 63}, 103505 (2001);  A. Mazumdar Nucl. Phys. B {\bf 597}, 561 (2001);
R. M. Hawkins and J. E. Lidsey, Phys. Rev. D  {\bf 63}, 041301 (2001);
S. C. Davis, W. B. Perkins, A.-C. Davis and I.R. Vernon,
Phys. Rev. D {\bf 63},  083518 (2001).
\bibitem{Bcmgy3}
E. J. Copeland, A. R. Liddle and J. E. Lidsey,
Phys. Rev. D  {\bf 64}, 023509 (2001);
D. Langlois, R. Maartens, and D. Wands,
      Phys. Lett. B {\bf489}, 259 (2000) ;
G. Huey and J. E. Lidsey, Phys. Lett. B {\bf 514}, 217 (2001) ;
V. Sahni, M. Sami, and T. Souradeep, gr-qc/0105121:
A. R. Liddle, and N. Taylor, astro-ph/0109412;
:A. S. Majumdar , Phys. Rev. D  {\bf 64}, 083503 (2001).


\bibitem{maeda}
K. Maeda,  Phys. Rev. D {\bf 64}, 123525 (2001).
\bibitem{mizuno_maeda}
S. Mizuno, and K. Maeda, Phys. Rev. D {\bf 64}, 123521  (2001).

\bibitem{Reheating}
Y. Shtanov, J. Traschen, and R. Brandenberger,
Phys. Rev. D {\bf 51}, 5438 (1995).

\bibitem{tsujikawa}
S. Tsujikawa, K. Maeda and S. Mizuno, Phys. Rev. D {\bf 63}
   ,123511 (2001).

\bibitem{maeda_wands}
K. Maeda, and D. Wands, Phys. Rev. D {\bf 62},  124009 (2000).


\bibitem{footnote1}
It is worth noting that in the
conventional cosmology, the power-law inflationary solution is obtained
in the model with an exponential potential, i.e.
$a\propto
t^p$
with
$p=2/\lambda^2$
for the
model
with $V(\phi)=\mu^4\exp[-\lambda\frac{\phi}{m_4}]$\cite{PL-inflation}).

\bibitem{PL-inflation}
F. Lucchin, and S. Matarrese, Phys. Rev. D {\bf 32},  1316 (1985).

\bibitem{maart_et.al}
R. Maatens, D. Wands, B. A. Bassett and I. P. C. Heard, Phys. Rev. D
{\bf 62}, 041301 (2000);


\bibitem{g.p.prod.}
L. H. Ford, Phys. Rev. D {\bf 35}, 2955 (1987);
L. P. Grishchuk, and Y. V. Sidrov,
Phys. Rev. D {\bf 42}, 3413 (1990);
B. Spokoiny, Phys. Lett. B {\bf315}, 40 (1993).

\bibitem{cope_et.al}
E. J. Copeland, A. R. Liddle and
D. Wands, Phys. Rev. D {\bf57}, 4686 (1998).


\end{thebibliography}
\end{document}